\documentclass[fleqn,usenatbib]{mnras}

\usepackage{newtxtext,newtxmath}
\usepackage{csvsimple}
\usepackage{booktabs}
\usepackage{adjustbox}
\usepackage{arydshln}

\usepackage[T1]{fontenc}

\DeclareRobustCommand{\VAN}[3]{#2}
\let\VANthebibliography\thebibliography
\def\thebibliography{\DeclareRobustCommand{\VAN}[3]{##3}\VANthebibliography}

\usepackage{graphicx}	
\usepackage{amsmath}	
\defcitealias{Quintana2025}{Q25}

\title[A new Gaia census of OB associations]{A new Gaia census of OB associations within 1 kpc}

\author[A. L. Quintana, et al. ]{Alexis L. Quintana$^{1}$\thanks{E-mail: alexis.quintana@obspm.fr}, Nicholas J. Wright$^{2}$\thanks{E-mail: n.j.wright@keele.ac.uk}, Lilly A. Kormann$^{3}$, João Alves$^{3}$,  David Katz$^{1}$,  \newauthor  Laia Casamiquela$^{1}$, Paola Di Matteo$^{1}$, Misha Haywood$^{1}$, and Chervin Laporte$^{1,4,5}$  \\
$^{1}$LIRA, Observatoire de Paris, Université PSL, Sorbonne Université, Université Paris Cité, CY Cergy Paris Université, CNRS, 92190 Meudon, France\\
$^{2}$Astrophysics Group, Keele University, Keele ST5 5BG, UK\\
$^{3}$University of Vienna, Department of Astrophysics, Türkenschanzstrasse 17, 1180 Wien, Austria\\
$^{4}$Institut de Ciències del Cosmos (ICCUB), Universitat de Barcelona, Martí i Franquès 1, E-08028 Barcelona, Spain \\
$^{5}$Kavli IPMU (WPI), UTIAS, The University of Tokyo, Kashiwa, Chiba 277-8583, Japan \\}

\date{Accepted 2026 April 29. Received 2026 April 29; in original form 2025 December 05 }

\pubyear{\the\year{}}

\begin{document}
\label{firstpage}
\pagerange{\pageref{firstpage}--\pageref{lastpage}}
\maketitle

\begin{abstract}
OB associations are primordial tracers of star formation and Galactic structure. Originally defined about 80 years ago, their historical membership lists have been superseded thanks to the precise astrometry from ESA's \textit{Gaia}'s satellite. Recent studies have however been mostly focused on individual OB associations or limited by the coverage of spectroscopic surveys. In this paper, we exploit a complete census of $\sim$25,000 O- and B-type stars within 1 kpc of the Sun to produce a highly-reliable catalogue of 56 OB associations using the HDBSCAN clustering algorithm, increasing the number of known OB associations by a factor of two within this volume. We assess the validity of this catalogue by crossmatching our OB association members with other catalogues of OB associations, star clusters and young stellar groups, confirming the high-confidence of our census of OB associations. We characterize these OB associations physically (total initial stellar mass, number of OB stars, ...) and kinematically (velocity dispersion, linear expansion ages, ...). The majority of the OB associations (38 out of 56) exhibit a significant expansion pattern in at least one direction, including 12 in both plane-of-the-sky directions, though differences in expansion velocity suggest anisotropical expansion patterns. We compare the locations of these OB associations with superclouds and features in the local Milky Way such as the Radcliffe Wave and discuss the implications for star formation in the solar neighbourhood.
\end{abstract}




\section{Introduction}
O- and B-type stars are the most massive and hottest stars according to the Harvard spectral classification \citep{Payne1925}. It is common to refer to massive stars as OB stars, with a metallicity-dependent minimum threshold of $\gtrsim$ 8--12 M$_{\odot}$ \citep[e.g.,][]{Jones2013,IbelingHeger2013}. These stars bring consequential amounts of feedback into the interstellar medium  (ISM, e.g. \citealt{Hopkins2014, Krumolhz2014,Dale2015}): in particular, the core-collapse supernova explosions following their deaths enrich the ISM with heavy elements that are used to create new stellar and planetary systems \citep[e.g.,][]{DeRossi2010}. Furthermore, due to their short life, OB stars constitute useful tracers of star formation \citep[e.g.,][]{Crowther2012}, as well as of the location of the Milky Way spiral arms \citep[e.g.,][]{Sparke2000}. As such, the distribution of OB stars is often correlated with other tracers of Galactic structure such as giant molecular clouds (GMCs),  H{\sc ii} regions and interstellar dust \citep[e.g.,][]{Chen2019,Zari2021,Vergely2022,PantaleoniGonzalez2025}.

OB stars are also the most prominent members of OB associations  \citep[e.g.,][]{Morgan1953,Humphreys1978,McKeeWilliams1997}. Originally defined by \citet{Ambartsumian1947} as low-density ($<$ 0.1 M$_{\odot}$ pc$^{-3}$) and gravitationally unbound stellar groups recognizable by their bright OB members, OB associations typically disperse within a few tens of Myrs, and therefore serve as key probes of early stellar evolution as a transitional phase between star-forming regions and the Galactic field population of stars \citep{Brown1997,Wright2020,Negueruela2025}. Because the distance of stellar groups is easier to determine than individual stars \citep[e.g.,][]{QuintanaWright2022}, OB associations constitute useful tracers of Milky Way structure as well \citep[e.g.,][]{Quintana2023}. Nevertheless, many historical (i.e, pre-\textit{Gaia}) catalogues of OB associations \citep[e.g.,][]{Humphreys1978,GarmanyStencel1992} have been defined only by the photometry or spectroscopy of their brightest members, and have been found to lack the kinematic coherence expected for young co-moving groups of stars \citep[e.g.,][]{Wright2020,Quintana,Quintana2023}, although some progress was enabled with HIPPARCOS astrometry \citep[e.g.,][]{HipparcosOBAssociations,BouyAlves2015}.

\textit{Gaia} data \citep{Gaia}, and particularly its last data release DR3 \citep{GaiaDR3}, which contains $\sim$1.5 billion sources with astrometry, offers an opportunity to create a new census of OB associations that are more kinematically-coherent than their historical counterparts, and therefore more likely to be real. \textit{Gaia} data has so far been mostly used to perform studies on individual OB associations \citep[e.g.,][]{WrightMamajek2018,Cantat2019,Armstrong2020,Squicciarini2021,MiretRoig2022,Szilagyi2023,SaltovetsMcSwain2024,Posch2025}. While there were also recent studies exploring the entire sky that have attempted to identify new systems, they were either based on spectroscopic catalogues \citep{Chemel2022,Liu2025}, or were limited to groups containing O-type stars \citep{MaizApellaniz2020,MaizApellaniz2022,Villafranca}. 

\citet[][hereafter Q25]{Quintana2025} produced a new and high-completeness census of $\sim$25,000 O- and B-type stars (T$_{\rm eff} >$ 10,000 K, hereafter OB stars) within the nearest 1 kpc, from which they derived a local star formation rate of $2896^{+417}_{-1}$ M$_{\odot}$ Myr$^{-1}$ and a local core-collapse supernova rate of $\sim$15--30 per Myr, that they extrapolated to a Galactic star formation rate of $0.67^{+0.09}_{-0.01}$ M$_{\odot}$ yr$^{-1}$ and a core-collapse supernova rate of 0.4--0.5 per century. As a follow-up of this study, our aim is to exploit this catalogue to identify OB associations within 1 kpc from the Sun, taking advantage of the precision of \textit{Gaia} data as well as modern clustering algorithms. This census will not only allow us to trace star formation within the solar neighborhood, but also the structure of the local disk. It is particularly relevant with the recent discoveries of new nearby structures, such as the Radcliffe Wave \citep{Alves2020}, the Split \citep{Lallement2019} and other superclouds \citep{Kormann2026}, which are traced by young stellar populations \citep[e.g.,][]{Tu2022,Kerr2023,Konietzka2024,Bobylev2025}. In particular, it was found that larger OB associations tended to be less likely to be associated with interstellar material, supporting the view that they were expanding \citep{Blaauw1964}; therefore, these superclouds can serve as indicators of the ages of OB associations.

This paper is structured as follows. In Section \ref{identification} we describe the process to identify kinematically-coherent groups from our sample of OB stars, and how we filter them to obtain a highly-reliable census of OB associations within 1 kpc. In Section \ref{analysis} we analyse physically and kinematically these OB associations, and notably study their expansion. In Section \ref{comp} we compare their memberships with individual and general OB associations alongside star clusters and young stellar groups. In Section \ref{structure} we compare them with nearby superclouds such as the Radcliffe Wave and the Split and discuss the implications for star formation in the local Milky Way. Finally, we conclude in Section \ref{conclusions}.

\section{Identification of OB associations within 1 kpc}
\label{identification}

In this section, we detail how we identify a reliable list of OB associations within 1 kpc. This first requires us to apply a clustering algorithm to our catalogue to identify groups (Section \ref{clustering}) with a careful selection of inputs (justified in Appendix \ref{parameter_space}).  We use simulations of the contamination rate of our sample (Appendix \ref{filtering_kinematical}) and its completeness (Appendix \ref{false_negatives}) to perform a final rejection of groups too extended in positional or velocity space to be genuine OB associations (Section \ref{filtering_extension}), before visualizing them and establishing nomenclature (Section \ref{nomenclature}). 

\subsection{Initial clustering with HDBSCAN}
\label{clustering}

HDBSCAN \citep{HDBSCAN} is a suitable clustering algorithm to identify new OB associations, specifically applied to \textit{Gaia} data \citep[e.g.,][]{SantosSilva2021,Chemel2022,Quintana2023}. Its ability to find groups at various densities gives it a significant flexibility over other similar algorithms such as DBSCAN \citep{DBSCAN}, though this can be at the cost of a high false positive rate, that requires users to apply careful filtering criteria.

For this work, we follow the approach of \citet{Kerr2023} and perform a 5D clustering on the Galactic Cartesian coordinates $X$, $Y$, $Z$ and the Galactic transverse velocities $c * V_l$ and $c * V_b$, where $c$ is a constant set to equalize the velocity and space components. The transverse velocities have been corrected from their motions compared with the Local Standard of Rest (LSR) with the values of (U,V,W)$_{\odot}$ = ($11.1^{+0.69}_{-0.75}$, $12.24^{+0.47}_{-0.47}$, $7.25^{+0.37}_{-0.36})$ km s$^{-1}$ from \citet{Schonrich2010}\footnote{We have adopted the value from \citet{Schonrich2010} even though there have been more recent, post-\textit{Gaia} estimates, because it is still widely used. The Gaia-derived value of $V_{\odot} = 8.0 \pm 8.4$ km s$^{-1}$ from \citet{Bovy2020} has large uncertainties, but does agree with the value from \citet{Schonrich2010}.}, following the approach from \citet{Szilagyi2023}. 

While clustering with HDBSCAN produces fewer variations compared with DBSCAN \citep{DBSCAN}, as it removes density as a free parameter, the resulting, identified groups still depend on the value of the chosen inputs. For this reason, we have ensured to select a combination of HDBSCAN parameters resulting in stellar groups that correspond to the physical definition of OB associations. The details of this process are provided in Appendix \ref{parameter_space}.

We have found the best configuration to be with values of \texttt{min\_cluster\_size} of 15, \texttt{cluster\_selection\_epsilon} of 30 and $c$ of 6 pc km$^{-1}$ s$^{-1}$, producing an initial list of 102 groups shared between 3859 OB members.

\subsection{Final selection of groups}
\label{filtering_extension}

We have reduced the number of candidate OB associations initially identified in Section \ref{clustering} from 102 to 62 based on a two-step process. Firstly, we have estimated a probability for each group through bootstrapping and chosen the highly-reliable threshold based on a simulation of false positive groups that could be potentially detected from a randomized distribution of OB stars. Secondly, we have estimated our incompleteness by calculating the probability to recover a test association placed on a randomized field of OB stars. In doing so we added back the small, nearby groups with a low observed probability back into our list of candidate OB associations. The full details of these simulations are provided in Appendix \ref{contaminants_incompleteness}.

Our final step was to discard any groups with positional or velocity dispersions too high to be genuine OB associations. To help select the threshold for this, we turned to the catalogue of OB associations from \citet{HipparcosOBAssociations}, which includes 12 nearby OB associations identified using HIPPARCOS astrometry. We looked into the equivalent OB associations in our sample (Sco-Cen, Per OB2, Per OB3, Vela OB2, Lac OB1...) and calculated their intrinsic dispersions, which they typically range 10--50 pc in position and 1--4 km s$^{-1}$. Nevertheless, contrasting with the list of Auriga OB associations identified with \textit{Gaia} DR3 from \citet{Quintana2023}, the new Aur OB1 and Aur OB2 are characterized by intrinsic positional dispersions in the range 50--100 pc in some dimensions, implying that some OB associations can retain some kinematic coherence despite being more extended in positional space. Based on these values, we have compromised between the liberal and conservative thresholds adopted in Appendix \ref{parameter_space} choose a threshold on 1D positional ($XYZ$) dispersion of 100 pc and on 1D intrinsic ($V_{l,b}$) dispersion of 4 km s$^{-1}$. 



Discarding any system with a position or velocity dispersion larger than these thresholds, we obtained a final list of 56 OB associations within 1 kpc, with a total of 2551 members\footnote{We stress that this is a final list of high-confidence OB associations. The 46 remaining groups could also correspond to genuine OB associations, which could have a more filamentary/elongated structure, or would need more precise data to be confirmed. We will also upload these rejected candidate OB associations to the CDS.}, which increases the census of OB associations within 1 kpc by a factor of 2 \citep{Wright2020}. 

We note that only $\sim$10 \% of the O- and B-type stars in the sample used belong to one of these OB associations, although this increases to $\sim$16 \% if we count the 3859 stars that were initially clustered. The local surface density star formation rate ($\sum_{\rm SFR}$) derived from the OB star catalogue used was compared with the value from the star clusters catalogue from \citet{HuntReffert2023,HuntReffert2024} in \citet{QuintanaHuntParul2025}, where they concluded that a majority of stars arise from clustered environments in the local Milky Way. Just like \citet{Quintana2023}, the explanation for this is three-fold. Firstly, our sample of OB stars is dominated by late B-type stars whose lifetime can reach hundreds of Myrs \citep{Ekstrom}, longer than the typical dissolution time of stellar groups, and therefore these late B-type stars could have been born in a clustered environment before becoming field stars \citep{LadaLada2003}. Secondly, we required a minimum of 15 OB stars to define an OB association, whereas some stellar groups (such as open clusters, moving groups and T associations) could include only a handful of OB stars. Thirdly, some of these OB stars could be runaways that were ejected from their natal clusters or association: Mart\'inez Garc\'ia et al. (submitted) recently found that 3.5--6 \% of the 24,487 B-type stars identified in \citetalias{Quintana2025} are probable runaways.

\subsection{Visualization and nomenclature}
\label{nomenclature}



\begin{figure*}
    \centering
    \includegraphics[scale = 0.6]{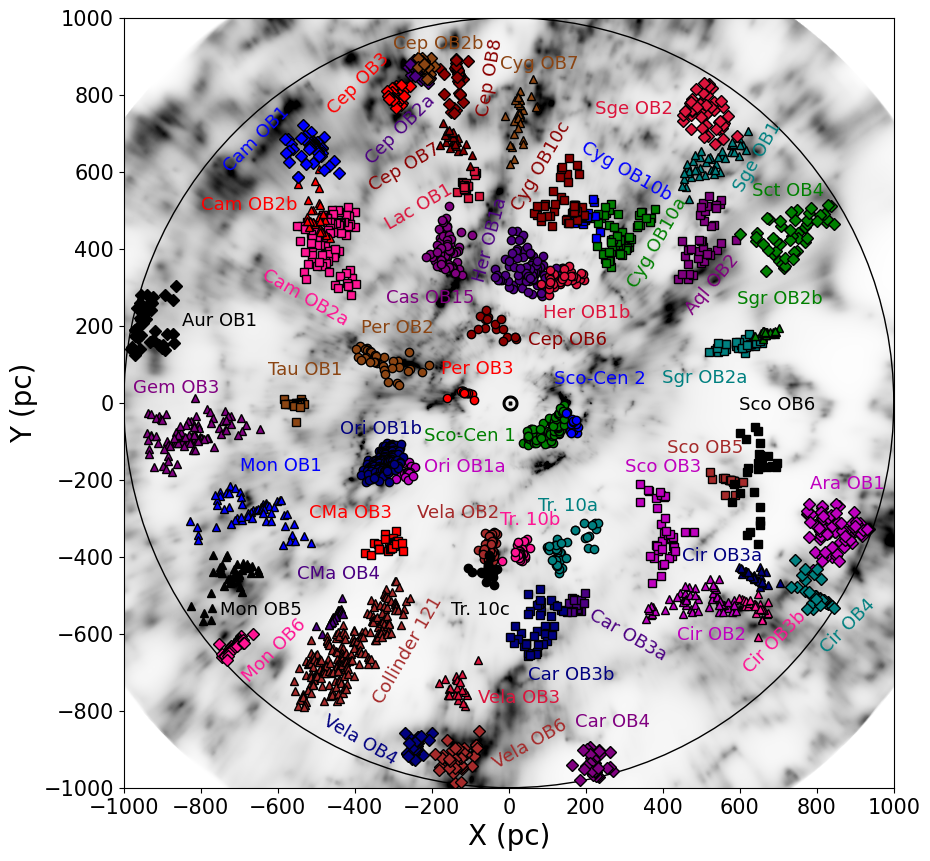}
    \includegraphics[scale = 0.6]{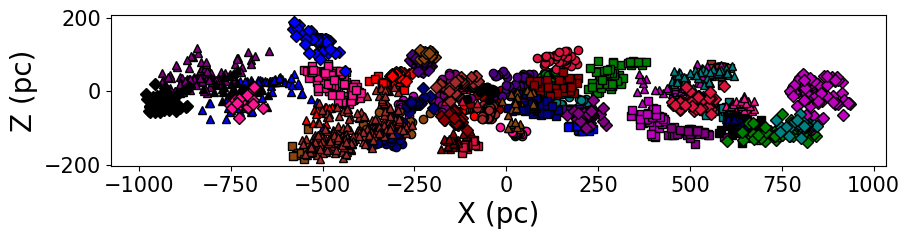}
    \includegraphics[scale = 0.6]{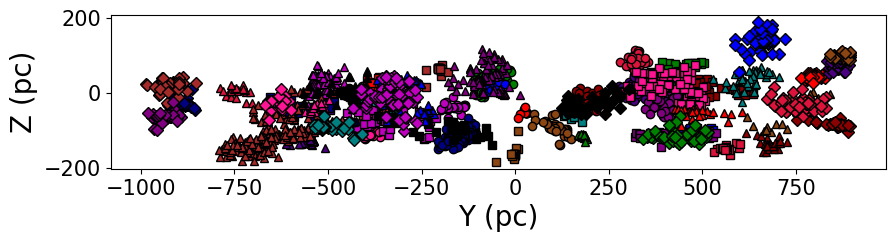}
    \caption{Galactic Cartesian coordinates of the 56 new OB associations, centered on the Sun's position. On the top panel have been labeled each OB association, displayed on top of the background extinction map from \citet{Edenhofer2024}. The colours used for each association are the same between panels.}
    \label{Map_OBAssociations}
\end{figure*}

Wa have plotted the 56 OB associations in Galactic Cartesian coordinates (X-Y plane) in Fig. \ref{Map_OBAssociations}, with their name indicated. The following conventions were used:

\begin{enumerate}
    \item We have followed the IAU convention for the nomenclature \citep{Ruprecht1966}, such that every OB association is named according to the constellation they belong to, followed by "OB", itself followed by a number. \footnote{The exceptions are Sco-Cen, Trumpler 10 and Collinder 121, because while the former's official name is Sco OB2, these OB associations have been constantly referred with these names over the past decades and still nowadays (e.g. \citealt{Wright2020}), therefore we prefer not to attribute them a new name to avoid confusion.}
    \item We compared the location of our new OB associations with historical censuses from \citet{HipparcosOBAssociations}, \citet{MelnikDambis2020} (who exploited the catalogue from \citealt{BlahaHumphreys1989} for their analysis) and \citet{Wright2020} (a review compiling the lists of OB associations from \citealt{Ruprecht1966}, \citealt{HipparcosOBAssociations} and \citealt{Brown1999}), and checked whether there was a significant overlap. This was notably the case for the most studied and well-known OB associations such as Sco-Cen, Ori OB1, Vela OB2 and Trumpler 10. If this was the case we adopted the classical name for these associations.
    \item When no counterpart with existing OB associations could be identified, we followed the historical convention of naming OB associations by their constellation (e.g. Aql OB2, Sct OB4, Sge OB1 and Sge OB2)\footnote{The exception to this was Gem OB3. \citet{Khachikian2005} identified several OB associations in the direction of the Gemini constellation, but the nomenclature was never used again, and only Gem OB2 was more recently confirmed as an actual OB association \citep{Khalatyan2024}.}.
    \item Finally, we consider our denomination to supersede the old ones: for instance, Ori OB1 has been historically divided in four subgroups (from a to d), which have been replaced here by the two associations Ori OB1a and Ori OB1b (ordered by increasing median distance) that do not correspond to the previous divisions.
\end{enumerate}

\begin{figure*}
    \centering
    \includegraphics[scale = 0.5]{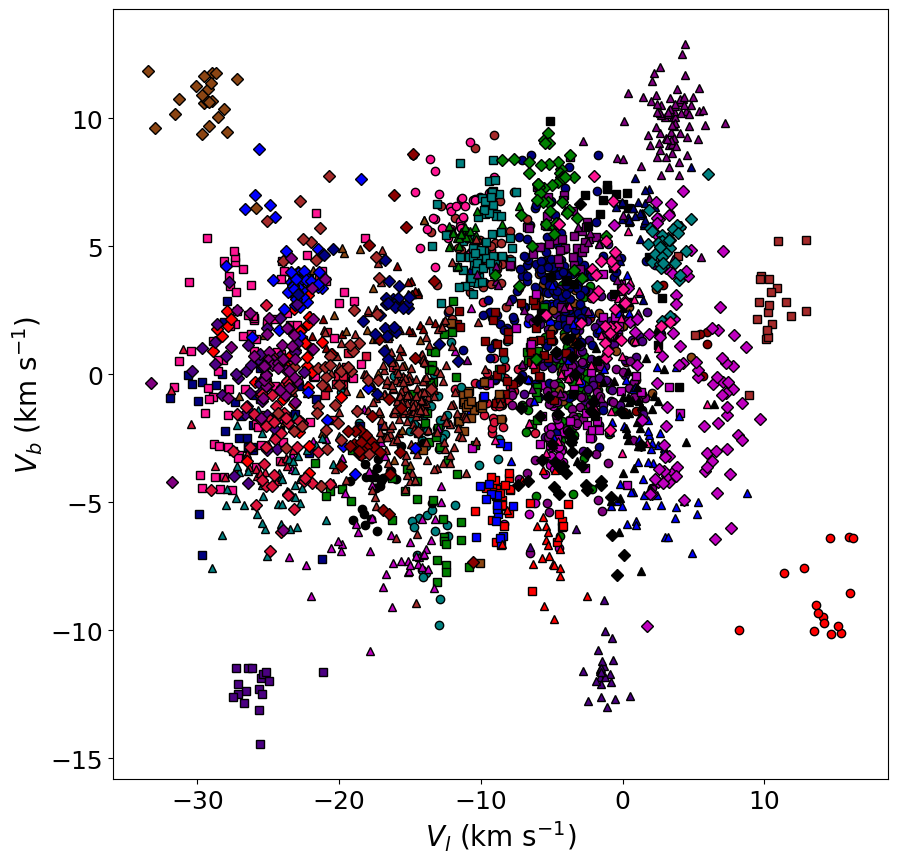}
    \caption{Transverse velocity distribution of the 56 new OB associations within 1 kpc. The colours used for each association are the same as in Fig. \ref{Map_OBAssociations}.}
    \label{VelAssoc}
\end{figure*}

In Fig. \ref{Map_OBAssociations} we also show the elevation of each OB association with respect to the Galactic plane. The scale height of our general OB star population is 76 $\pm$ 1 pc \citepalias{Quintana2025}, it is therefore not surprising to see that most of our OB associations sit close to the Galactic Plane, with the lowest and highest ones being respectively Tau OB1 (at Z = -166 pc) and Cam OB1 (at Z = 138 pc). 

The distribution of the OB associations in transverse velocities is shown in Fig. \ref{VelAssoc}. Unveiled here is a picture of kinematic coherence consistent with our clustering method. Most OB associations occupy the same broad region of velocity space, although groups like Car OB3a, CMa OB4, Per OB3, Gem OB3 and Cep OB2b stand out as outliers.

\section{Analysis of the new OB associations}
\label{analysis}

\begin{figure*}
    \centering
    \includegraphics[scale = 0.28]{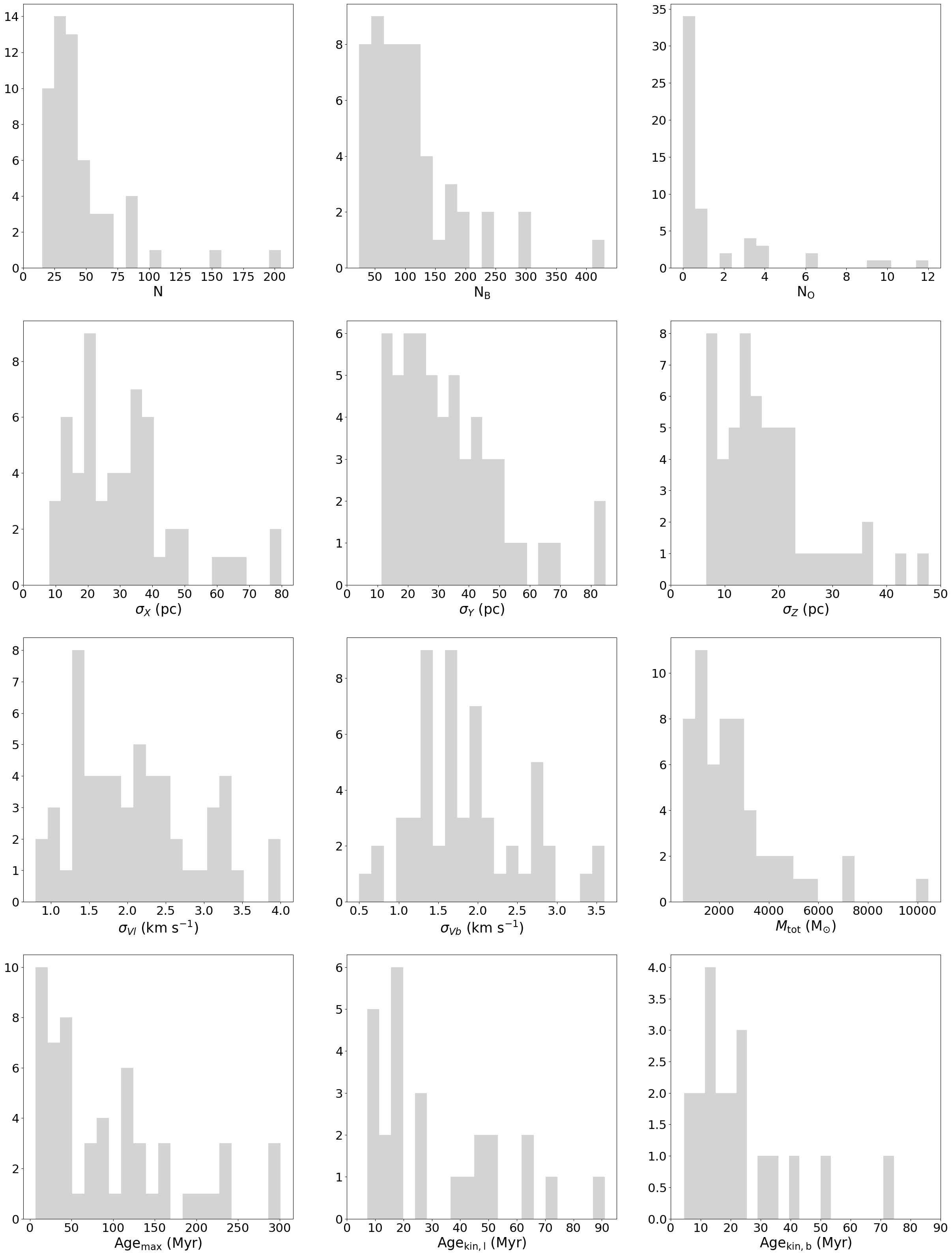}
    \caption{Distribution of the properties of our 56 OB associations derived in Section \ref{analysis} and described in Table \ref{Table_OBAssociations}. From left to right and from top to bottom, the parameters are respectively: number of identified members, number of systems with a B- and O-type star, intrinsic positional dispersions (in $X$, $Y$ and $Z$), intrinsic velocity dispersions (in $V_l$ and $V_b$), total initial stellar mass, maximum isochronal ages and kinematic ages.}
    \label{Histogram_Parameters}
\end{figure*}

\begin{figure*}
    \centering
    \includegraphics[scale = 0.23]{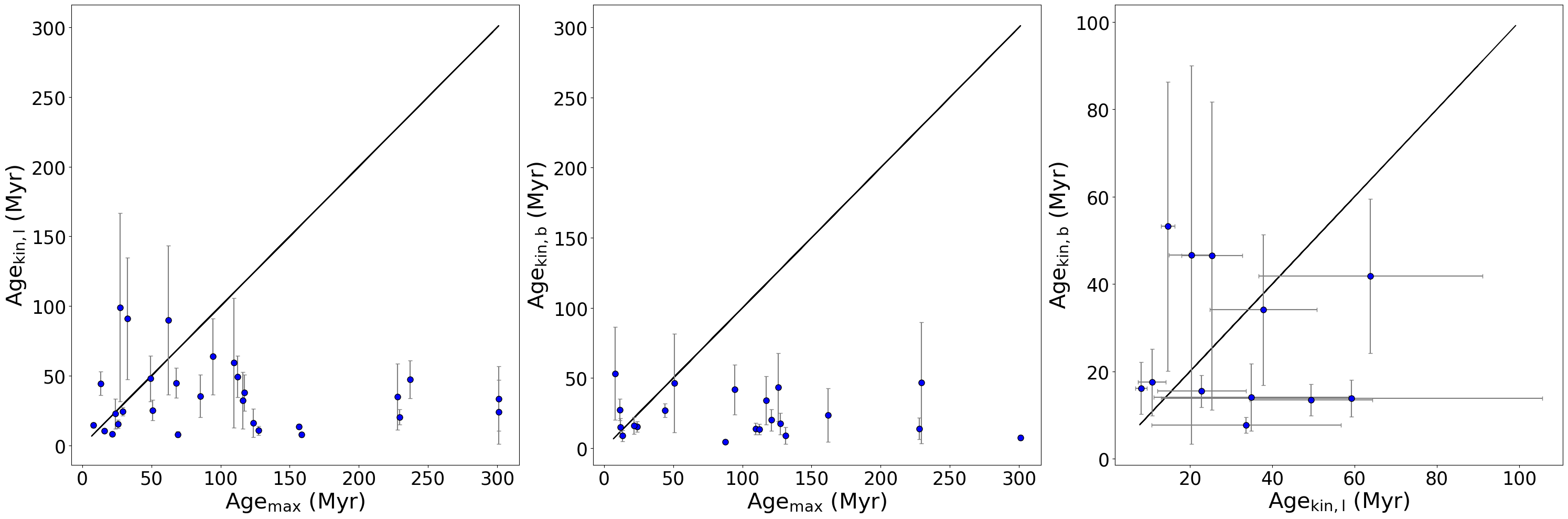}
    \caption{Comparison between the different age estimates for OB associations as derived and described in Sections \ref{maxages} and \ref{linearkinematicages}.}
    \label{Comp_Ages}
\end{figure*}

This section is dedicated to the physical and kinematical analysis of our 56 OB associations. The parameters we have derived for each of them are presented in a table published electronically and available at the Vizier archive, whose column names and descriptions are presented in Table \ref{Table_OBAssociations}, and illustrated with Ori OB1b as an example. The distributions of the parameters described in Table \ref{Table_OBAssociations} are also displayed as histograms in Fig. \ref{Histogram_Parameters}. In addition, we have shown the comparison between the maximum isochronal ages and the kinematic ages derived in Fig. \ref{Comp_Ages}, although little correlation is visible due to the limitations of each method, as outlined respectively in Sections \ref{maxages} and \ref{linearkinematicages}.

\begin{table*}
  \centering
  \renewcommand{\arraystretch}{1.2}
    \caption{Properties of the 56 OB associations derived in this work with their description, illustrated with Ori OB1b. The corresponding table is accessible on Vizier. \label{Table_OBAssociations}}
\begin{tabular}{lcll} 
\hline
Column label & Units & Description & Example \\  
\hline
Name & -  & Assigned OB association name & Ori OB1b \\
Alternative name & - & Alternative name for the OB association & - \\
ID & - & ID assigned to the OB association, ordered by increasing median line-of-sight distance & 9 \\
N & - & Number of identified members & 156 \\
RA\_ICRS & deg & Median right ascension (ICRS) at Ep=2016.0 & 83.61 \\
DE\_ICRS & deg & Median declination (ICRS) at Ep=2016.0 & 1.69 \\
GLON & deg & Median Galactic longitude & 205.32 \\
GLAT & deg & Median Galactic latitude & 18.33 \\
AV & mag & Median extinction in the V-band & 0.1 \\
d & pc & Median SED-fitted line-of-sight distance & 373.21 \\
X & pc & Median X coordinate in heliocentric Cartesian Galactic coordinates & -317.81 \\ 
Sigma\_X & pc & Intrinsic dispersion in the X direction & 35.2 \\
Y & pc & Median Y coordinate in heliocentric Cartesian Galactic coordinates & -154.86 \\ 
Sigma\_Y & pc & Intrinsic dispersion in the Y direction & 31.8 \\
Z & pc & Median Z coordinate in heliocentric Cartesian Galactic coordinates & -117.74 \\ 
Sigma\_Z & pc & Intrinsic dispersion in the Z direction  & 23.4 \\
V\_l & km s$^{-1}$ & Median transverse velocity in Galactic longitude & -4.93 \\
Sigma\_Vl & km s$^{-1}$ & Intrinsic dispersion in transverse velocity in Galactic longitude & 2.5 \\
V\_b & km s$^{-1}$ & Median transverse velocity in Galactic latitude & 3.56 \\
Sigma\_Vb & km s$^{-1}$ & Intrinsic dispersion in transverse velocity in Galactic latitude & 2.0 \\
Kappa\_l  & km s$^{-1}$ pc$^{-1}$ & Linear velocity gradient in Galactic longitude & $0.069 \pm 0.008$ \\
Age\_kin,l & Myr & Linear kinematic age in Galactic longitude & $14.6 \pm 1.7$ \\
Kappa\_b & km s$^{-1}$ pc$^{-1}$ & Linear velocity gradient in Galactic latitude & $0.019 \pm 0.012$ \\
Age\_kin,b & Myr & Linear kinematic age in Galactic latitude & $53.2 \pm 33.1$ \\
N\_O & - & Estimated number of systems containing an O-type star in the OB association & $12^{+7}_{-0}$ \\
N\_B & - & Estimated number of systems containing a B-type star in the OB association  & $296^{+9}_{-65}$ \\
M\_tot & Msun & Total stellar initial mass & $7252^{+507}_{-1403}$ \\
Age\_max & Myr & Estimated maximum isochronal age from the hottest member & 8.0 \\
N\_OC & - & Number of crossmatched members with the OCs from \citet{HuntReffert2024} & 31 \\
Related OCs age & Myr & Age range of the overlapping OCs from \citet{HuntReffert2024} & 4--13 \\
Related OCs & - & Crossmatched OCs from \citet{HuntReffert2024} & See Table \ref{TableOC} \\
N\_G & - & Number of crossmatched members with the stellar groups from \citet{Kounkel2020} & 119 \\
Related groups age & Myr & Age range of the overlapping stellar groups from \citet{Kounkel2020} & 5 \\
Related groups & - & Crossmatched groups from \citet{Kounkel2020} & Orion \\
Associated supercloud & - & Spatially related supercloud from \citet{Kormann2026} & Radcliffe Wave \\

\hline
\end{tabular}
\end{table*}

\subsection{Physical properties of the new OB associations}
\label{physical}

\subsubsection{Number of O- and B-type stars}
\label{number}

We have followed a similar method to \citet{Quintana} and \citet{Quintana2023} to estimate the total number of O- and B-type stars in each OB association. B-type stars are defined as those with $4 < \log(T_{\rm eff}) < 4.3$, whereas O-type stars as those with $\log(T_{\rm eff}) > 4.3$ (using the values estimated from the spectral energy distribution, i.e. SED, fitting process in \citetalias{Quintana2025}), akin to the approach from Mart\'inez Garc\'ia et al. (submitted). Because of the high multiplicity of O- and B-ype stars \citep{DucheneKraus2013}, and because the SED fitter applied in \citetalias{Quintana2025} does not resolve binarity, we will refer to these quantities as "number of systems containing O/B-type stars" afterwards.

However, compared with \citet{Quintana} and \citet{Quintana2023}, the effective temperature threshold we chose for our OB stars in \citetalias{Quintana2025} (i.e., T$_{\rm eff}$ $>$ 10,000 K) means that we are not including cooler evolved massive stars in our census of OB association members. Therefore, to account for them, we use the SED-fitted mass of each OB association member to interpolate the stellar effective temperature as a function of their age from the stellar evolutionary models from \citet{Ekstrom}. This allows us to estimate which fraction of their lifetime a star of a given initial mass spends with $\log(T_{\rm eff}) < 4$: thus, each time we add a star to the total number, we divided by this fraction to account for high-mass non-OB stars (in practice, this re-normalisation typically increased the number of stars by $\sim$10 \%). Uncertainties on the number of OB systems were estimated through a Monte Carlo (MC) simulation where the SED-fitted effective temperatures were randomly sampled within their uncertainties 1000 times. 

The results show a median value of $\sim$ 95 systems containing a B-type star in our OB associations. 41 out of the 56 OB associations are consistent with containing at least 1 system with an O-type star. Particularly noteworthy are Ori OB1b, Cep OB3, Collinder 121, Sco-Cen 1 and Vela OB6, which include $12^{+7}_{-0}$, $10^{+1}_{-2}$, $9^{+4}_{-3}$, $6^{+3}_{-0}$ and $6^{+4}_{-0}$ O-type stars, respectively. This picture is consistent with these OB associations being spatially correlated with their surrounding gas (see Fig. \ref{Map_Superclouds} and \ref{YZ_Superclouds}), with the notable exception of Collinder 121. Only Cep OB3 is however significantly reddened (A$_V$ $\sim$2.1 mag), whereas all these other OB associations have a median A$_V$ lower than 0.5 mag.

\subsubsection{Total initial stellar mass}
\label{stellarmass}

Our estimation of the total initial stellar mass for each OB association is based on the method used in Section 4.1 from \citetalias{Quintana2025} to derive the star formation rates from our census of OB stars. We used the number of OB stars estimated per OB association in Section \ref{number}, and generated the full stellar population within the group using the initial mass function from \citet{IMFMasch}, with masses sampled between between 0.01 and 100 M$_{\odot}$ and with a high-mass exponent of $\alpha = 2.3$. To account for unresolved binarity, we determined if each star belonged to a binary system randomly with the mass-dependent binary fractions from \citet{DucheneKraus2013}. Then, for binary stars, we have generated a random mass ratio $q =M_2/M_1$, (assuming a flat distribution between 0 and 1) to estimate the mass of the secondary star, neglecting binary interaction. The total stellar initial mass for each OB association corresponds to the sum of the individual stellar masses for either the single stars or both components of the binary systems, as the median value of a process repeated 1000 times, to account for the fluctuations of the initial mass function (IMF), and the randomness in drawing the values of the binary probability and fraction. The errors are drawn from a MC simulation with 1000 iterations, using the uncertainties on the estimated number of OB stars. 

The median total initial stellar mass of all our associations is $\sim$2200 M$_{\odot}$, consistent with the typical mass of OB associations \citep{Wright2020}. For specific OB associations, \citet{Wright2020} placed the value for Ori OB1 in the range 4000--13,000 M$_{\odot}$, also using the IMF from \citet{IMFMasch}, whilst \citet{Blaauw1964} and \citet{Briceno2019} stated it should be larger than 8000 M$_{\odot}$. If we sum the $M_{\rm tot}$ from our Ori OB1a and Ori OB1b, we obtain a total initial stellar mass of $\sim$9000 M$_{\odot}$, consistent with past estimates. Meanwhile, estimations of the total initial stellar mass of Vela OB2 vary, between $\sim$1300 M$_{\odot}$ \citep{Armstrong2018} and 2330 M$_{\odot}$ \citep{Cantat2019}, while our estimate is equal to$\sim$2360 M$_{\odot}$ and therefore at the upper limits of this interval. \citet{WrightMamajek2018} estimated a $M_{\rm tot}$ of $\sim$4000 M$_{\odot}$ from the individual estimates of \citet{Mamajek2002} and \citet{PreibischMamajek2008} for each of the historical Sco-Cen subgroups, slightly smaller than our value of $\sim$5000 M$_{\odot}$ for Sco-Cen 1 and Sco-Cen 2 combined. Finally, as expected because of our coverage limit, the total initial stellar mass from Aur OB1 ($\sim$2800 M$_{\odot}$) is lower than in \citet{Quintana2023} where it was estimated to be $\sim$6000 M$_{\odot}$, which was also likely under-estimated as the derivation method did not account for unresolved binarity.

\subsubsection{Maximum isochronal ages}
\label{maxages}

While deriving an isochronal age for our OB associations can be non-trivial, given that our membership only covers the upper-left part of the HR diagram, it is possible to estimate an upper age limit, based on the most massive members of each association. To that end, we have interpolated the hottest star from each OB association with the rotating stellar evolutionary models from \citet{Ekstrom}, as they correspond to the members with the shortest lifetime, allowing us to calculate a maximum age for each OB association. We however stress that these should be considered as approximate values, as the evolutionary models from \citet{Ekstrom} are valid for single stars and neglect binary interaction.

There are some instances where the maximum age is roughly consistent with literature values, such as Sco-Cen 1 with an age of $\sim$12 Myr (compared with 3--19 Myr in \citealt{MiretRoig2022,Ratzenbock,Posch2025}) , Ori OB1b  with an age of $\sim$8 Myr (compared with 1--12 Myr in \citealt{Bally2008,Briceno2019,Sanchez2024}), Tr. 10b with an age of $\sim$42 Myr (compared with 45--50 Myr in \citealt{Cantat2019,Kerr2023}), Per OB2  with $\sim$10 Myr (compared with 2--7 Myr in \citealt{Bally2008,Kounkel2022}), and Cyg OB7 aged $\sim$11 Myr (compared with $\sim$13 Myr in \citealt{Uyaniker2001}).

In many cases, however, the maximum ages are either slightly overestimated, such as Cep OB2 aged 21--24 Myr (compared with 3--16 Myr in \citealt{Szilagyi2023}) and Cep OB3 aged $\sim$13 Myr (compared with $\sim$3 Myr in \citealt{Kerr2023}), or even strongly overestimated as in the cases of Sco-Cen 2 aged $\sim$50 Myr, and Per OB3 aged $\sim$131 Myr (compared with $\sim$58 Myr in \citealt{Kerr2023}).

Some of these age disagreements, particularly for Vela OB2, could be due to temporal substructures in that association. Indeed, noteworthy is this association which has the lowest maximum age of $\sim$7 Myr, younger than the commonly-accepted age of the Vela OB2 complex. This could be because it includes $\gamma^2$ Vel, a famous Wolf-Rayet binary system, part of the $\gamma$ Velorum cluster and thought to have formed after the bulk of the lower-mass stars in Vela OB2 \citep{Jeffries2009}.

\subsection{Kinematic properties of the new OB associations}
\label{kinematic}

\subsubsection{Positional and velocity dispersion}

We have calculated both the positional and velocity dispersions of the OB associations. The median dispersion in $X$ (29.0 pc) and in $Y$ (30.1 pc) are similar, whilst the median dispersion in $Z$ is smaller (15.8 pc) because of the restoring effect of the Galactic disk reducing expansion in that direction. Likewise, OB associations typically share a similar extent in velocities in both directions (2.0 and 1.7 km s$^{-1}$, respectively): these are significantly smaller than the value of 4.5 km s$^{-1}$ derived for the 28 OB associations in \citet{MelnikDambis2020}, but their study relied on historical memberships of OB associations. 

Our 1D velocity dispersions in the Sco-Cen associations are in the range 1.0--1.5 km s$^{-1}$, which is compatible with the average values of $1.86 \pm 0.21$ km s$^{-1}$, $1.38 \pm 0.21$ km s$^{-1}$ and $1.21 \pm 0.28$ km s$^{-1}$ calculated by \citet{WrightMamajek2018} for Upper Scorpius (US), Upper Centaurus Lupus (UCL) and LCC, respectively. Estimates on the 1D velocity dispersions of Ori OB1 are typically within the range 1.5--3 km s$^{-1}$ depending on the region studied \citep[e.g.,][]{VanAltena2008,Furez2008,DaRio2017}, again compatible with our results (respectively 1.0--1.6 km s$^{-1}$ for Ori OB1a and 2.0--2.5 km s$^{-1}$ for Ori OB1b). Likewise, \citet{Jeffries2014} calculated a radial velocity (RV) dispersion of $1.60 \pm 0.37$ km s$^{-1}$ for the surrounding population of $\gamma$ Vel, which encompasses part of Vela OB2, and is consistent with our results for Vela OB2 (1.9--2.0 km s$^{-1}$ ). Finally, the RV dispersion of $\sim$3.9 km s$^{-1}$ estimated by \citet{Steenbrugge2003} for the early-type stars of Per OB2 is significantly higher than our velocity dispersions (1.7--2.7 km s$^{-1}$) for this association, but \citet{Wright2020} pointed out this value was likely overestimated as they did not account for unresolved binary motions in their calculations. 

\subsubsection{Kinematic ages}
\label{linearkinematicages}

OB associations typically dissolve after a few tens of Myrs \citep{Ambartsumian1949}, and recent investigations with \textit{Gaia}-based memberships have revealed significant expansion signatures within them \citep[e.g.,][]{Cantat2019,Quintana,Grobschedl2025}. Conversely, \citet{WardKruijssen2018} did not detect any expansion in their 18 studied OB associations, whereas \citet{MelnikDambis2020} only succeeded to do so in 6 out of 28. In both cases, however, their analysis relied upon historical (pre-\textit{Gaia}) censuses crossmatched with \textit{Gaia} data, while we identified members directly from \textit{Gaia} data. Our new census contains significantly more OB associations than \citet{WardKruijssen2018} and \citet{MelnikDambis2020}, and therefore represents an opportunity for a more statistically-significant analysis of expansion signatures in these groups. 

\begin{figure}
    \centering
    \includegraphics[scale = 0.35]{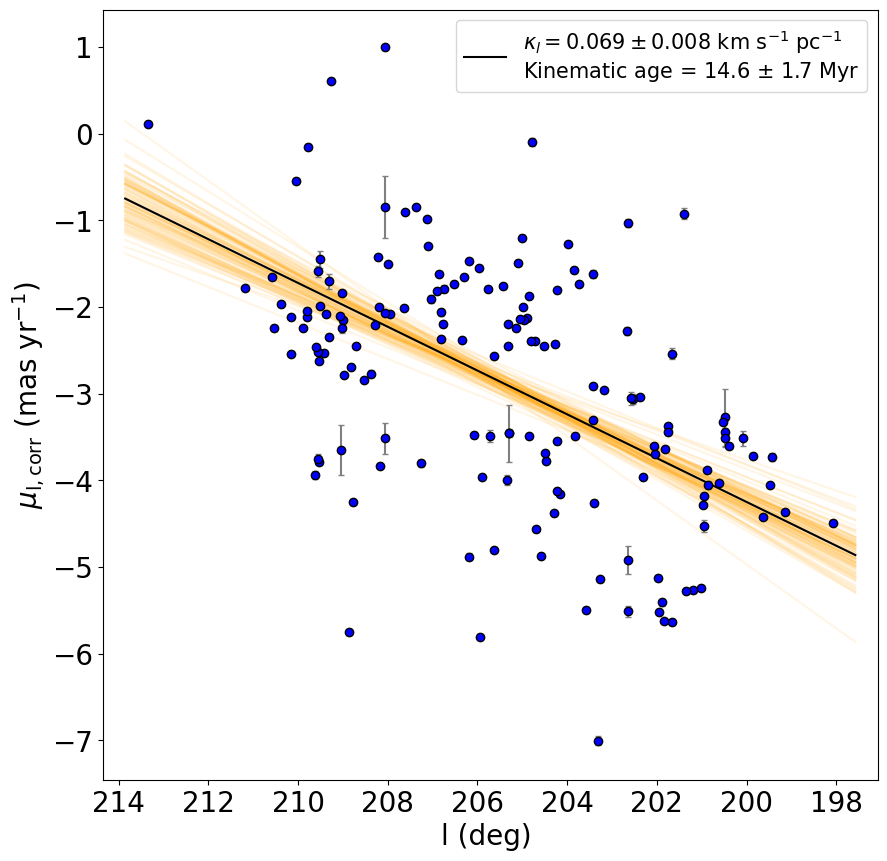}
    \caption{Fitted linear gradient (and the corresponding kinematic age) between the Galactic longitude and the proper motions in Galactic longitude (corrected for the LSR and virtual expansion) for the Ori OB1b members. 100 random samples have been drawn from the posterior distribution of the linear regression and have been displayed in light orange to illustrate the uncertainty on the fitted gradient.}
    \label{LinearFit_OriOB1b}
\end{figure}

Here we apply a linear fit between the Galactic coordinates of the OB association members and their proper motions in Galactic coordinates (corrected for the LSR as in Section \ref{clustering}, and also corrected for the projection effects, referred to as "virtual expansion", as detailed in Appendix \ref{correction_virtual}), along each axis, using a weighted least squares linear regression model,. In doing so we estimate velocity gradients in the $l$ and $b$ directions, that we convert from units of mas yr$^{-1}$ deg$^{-1}$ to units of km s$^{-1}$ pc$^{-1}$. If the OB association exhibits a significant expansion pattern (defined as having a gradient $\kappa >$ 0.01 km s$^{-1}$ pc$^{-1}$), we invert the value to obtain a kinematic age. An example of this linear fit is displayed in Fig. \ref{LinearFit_OriOB1b} for Ori OB1b, which exhibits a significant expansion pattern along the Galactic longitude direction.

We find that 38 out of our 56 OB associations, hence the majority of our catalogue ($\sim$68\%), exhibit a significant ($>1\sigma$) expansion pattern in at least one direction, to be compared with $\sim$94 \% of the 18 stellar groups in \citet{Wright2024} (that includes star clusters and OB associations) expanding in at least one direction. 

Only 12 of our OB associations exhibit expansion patterns in both directions (see right panel from Fig. \ref{Comp_Ages}), but kinematic ages are only consistent for Cep OB6 (10.7 $\pm$ 3.3 Myr in $l$ and 17.5 $\pm$ 7.6 Myr in $b$), Cam OB2a (63.9 $\pm$ 27.2 Myr in $l$ and 34.1 $\pm$ 17.3 Myr in $b$), Mon OB5 (37.9 $\pm$ 13.0 Myr in $l$ and 41.8 $\pm$ 17.6 Myr in $b$) and Cep OB2a (22.8 $\pm$ 10.8 Myr in $l$ and 15.5 $\pm$ 3.7 Myr in $b$). For other OB associations, they are either inconsistent (e.g. 49.4 $\pm$ 15.0 Myr in $l$ and 13.5 $\pm$ 3.6 Myr for Aur OB1), or not significant enough because of the large error bars in one direction (e.g. 14.6 $\pm$ 1.7 Myr in $l$ and 53.2 $\pm$ 33.1 Myr in $b$ for Ori OB1b and 25.3 $\pm$ 7.4 Myr in $l$ and 46.5 $\pm$ 35.2 Myr in $b$ for Sco OB3). Whether these differences can be attributed to large measurement uncertainties or actual asymmetrical expansion patterns warrant a more accurate method relying on 3D kinematics and accounting for the effects of Galactic potential in the association's expansion, as in \citet{QuintanaWright2022}.

 Our kinematic ages roughly align with literature ages for some OB associations such as Cas OB15 (15.3 $\pm$ 3.1 Myr in $l$ compared with 6--11 Myr for Alessi 20 in \citealt{Pang2022,Kerr2023,HuntReffert2024,ArmstrongTan2026}), Her OB1a (26.8 $\pm$ 4.9 Myr in $b$ compared with $\sim$29 Myr for Cep-Her in \citealt{Kerr2023}), Tr. 10a (44.9 $\pm$ 10.7 Myr compared with 45--50 Myr in \citealt{Cantat2019,Kerr2023}), Cir OB3b (7.9 $\pm$ 2.0 Myr in $b$ compared with 1--9 Myr for the subgroups of the Circinus complex in \citealt{Kerr2025}), Cep OB2 (5--20 Myr for both subgroups in both directions compared with 3--13 Myr in \citealt{Kerr2023,Szilagyi2023}) and Cep OB3 (9.0 $\pm$ 4.0 Myr in $b$ compared with $\sim$3 Myr in \citealt{Kerr2023}). Moreover, Per OB3, Cep OB6, Ori OB1b, Cas OB15, Cyg OB10a, Car OB3a, Car OB3b, Cep OB7, Sco OB3, CMa OB4, Cir OB3a, Cir OB3b, Gem OB3, Cep OB3, Vela OB6, Cam OB1, Cep OB8, Cep OB2a, Cep OB2b, Mon OB6, Car OB4 and Aur OB1 all have median kinematic ages younger than 20 Myr in at least one direction.

While this picture looks consistent with OB associations containing O-type stars (e.g. Cep OB3 and Vela OB6), the kinematic age of Ori OB1b  (see Fig. \ref{LinearFit_OriOB1b}) seems slightly over-estimated compared with literature estimates (1--12 Myr, \citealt{Bally2008,Briceno2019,Sanchez2024}). Likewise, we have estimated a kinematic age of 32.2 $\pm$ 20.3 Myr in $l$ for Lac OB1 (compared with an isochronal age of $\sim$10 Myr in \citealt{Kerr2023}), whilst the kinematic age estimated for Cir OB2 (24.4 $\pm$ 3.2 Myr in $l$) is noticeably larger than the ages of the subgroups (1--9 Myr) identified in the Circinus complex in \citet{Kerr2025}. Furthermore, we do not observe a linear expansion pattern in any OB association of the Sco-Cen complex, even though it is believed to be expanding (e.g. \citealt{Grobschedl2025}), and attribute this to the high number of OB members coming from the Bright Stars Catalogue and fitted with HIPPARCOS data because they were saturating in \textit{Gaia} (as explained in \citetalias{Quintana2025}), thus having larger astrometric uncertainties and preventing us from detecting significant linear expansion patterns.

Contrariwise, the age of Gem OB3 (13.7 $\pm$ 1.7 Myr in the $l$ direction) looks underestimated given it is the most extended OB association in the $X$ direction, while the kinematic ages for Cep OB6 (10.7 $\pm$ 3.3 Myr in $l$ and 17.5 $\pm$ 7.6 Myr in $b$) are significantly smaller than the isochronal age of $\sim$59 Myr derived by \citet{Kerr2023}.

This comparison stresses that this method only provides generally upper limits for the ages of OB associations, as discussed in \citet{Quintana}. OB associations have been found to be characterized by kinematic substructure \citep[e.g.,][]{Wright2016,Goldman2018,Zerjal2023}, and significant age spreads \citep[e.g.,][]{Wright2015,Kounkel2018} that make it harder to assign them a unique age. Some of the uncertainties we derived could reflect this observed age spread (e.g., the age of 35.5 $\pm$ 15.2 Myr for Cam OB1 in $l$) while others rather reflect the limitations of the method (e.g., 63.9 $\pm$ 27.2 Myr for Cam OB2a in $l$). The ratio between the uncertainties on the velocity gradient and the velocity gradient typically increases for decreasing number of identified members (most OB associations with N $<$ 40 have $\frac{\sigma_\kappa}{\kappa}$ > 1 have in both directions). 

Therefore, extended membership, alongside an accurate 3D kinematical traceback, will be necessary if we are to calculate more accurate kinematic ages in the future\footnote{We have not estimated 3D kinematic ages for the moment because, as stressed in Appendix \ref{correction_virtual}, only $\sim$54\% of OB association members have a valid RV measurement. Moreover, the majority of these RV measurements originate from \textit{Gaia} DR3, whose precisions are weaker for late B-type stars \citep{Blomme2023,Katz2023}, which may require us to expand the membership of our OB associations to lower-mass stars to derive reliable 3D kinematics.}. The internal kinematics of our OB associations will be investigated in more detail in a follow-up work (Olivares, Quintana, Miret-Roig and Bouy, in prep.), wherein we will derive more accurate kinematic ages.

\section{Comparison with other catalogues}
\label{comp}

In this section we compare our OB associations with external catalogues of OB associations, as well as other catalogues of stellar groups (e.g. open clusters), in order to contrast, analyse the level of overlap, validate our catalogue and relate it to young stellar structures and groups of the local Milky Way. 

For each catalogue, we have performed the crossmatch using the source IDs: for catalogues based on \textit{Gaia} DR3, we were able to perform this directly, while for catalogues based on \textit{Gaia} DR2, we used the \texttt{dr2\_neighborhood} table in the \textit{Gaia} archive, and for catalogues based on HIPPARCOS, we used the \texttt{astroquery} package from Python as in Section \ref{filtering_extension}. A comparison with catalogues of individual OB associations is also discussed in Appendix \ref{indiv_OBassoc}.

\begin{table*}
	\centering
	\caption{Comparison between our 2551 OB association members with external catalogues. N$_{\rm C}$ is the total number of stars in the external catalogue within $\sqrt{X^2+Y^2} <$ 1 kpc whereas N$_{\rm O}$ stands for our number of OB association members successfully crossmatched with the corresponding catalogue.}
    \label{CompCatalogues}
	\renewcommand{\arraystretch}{1.3} 
	\begin{tabular}{lccccccr} 
		\hline
		Catalogue & Type & Region & Data used & N$_{\rm C}$ & N$_{\rm O}$ \\
		\hline
        \citet{MelnikDambis2020} & OB associations & All-sky & \textit{Gaia} DR2 & 262 & 89 \\
        \citet{HipparcosOBAssociations} & OB associations & All-sky ($<$ 1 kpc) & HIPPARCOS & 1130 & 168 & \\
        \citet{BouyAlves2015} & OB groups & All-sky ($<$ 500 pc)  & HIPPARCOS & 123 & 23 \\
        \citet{Chemel2022} & OB associations and star clusters &  Galactic plane ($|b| \leq 20\degr$) & \textit{Gaia} DR3 & 1996 & 431  \\
        \citet{HuntReffert2024} & Star clusters & All-sky & \textit{Gaia} DR3 & 211,300 & 985 \\
        \citet{Kerr2023} & Young stellar groups & All-sky &  \textit{Gaia} DR3 & 36,182 & 313  \\
        \citet{Kounkel2020} & Star clusters, co-moving groups and structures & All-sky &  \textit{Gaia} DR2 & 351,984 & 1097  \\
		\hline
	\end{tabular}
\end{table*}

\subsection{General OB associations}
\label{compobassoc}

The membership of historical OB associations was originally compiled in \citet{Humphreys1978} and \citet{BlahaHumphreys1989}. The most recent analysis of this catalogue was conducted by \citet{MelnikDambis2020}, where they identified 1990 out of the 2209 historical OB associations containing \textit{Gaia} DR2 astrometry \citep{GaiaDR2}, including 262 within $\sqrt{X^2+Y^2} <$ 1 kpc. The results of our crossmatch with this catalogue are displayed in Table \ref{CompCatalogues}, with only 89 members in common. Nearly half of the overlap (39 out of 89) between our catalogue and theirs correspond to Ori OB1 members in \citet{MelnikDambis2020}, but we also recover 9 members from the historical Cep OB3, as well as 6 members from Vela OB2 and Collinder 121, 5 members from Cyg OB7 and Ara OB1a and 4 members from Per OB2. Out of the 173 remaining stars from \citet{MelnikDambis2020}, 90 successfully crossmatch with our catalogue of OB stars from \citetalias{Quintana2025}. This is unsurprising given that previous papers (e.g. \citealt{Quintana,Chemel2022,Quintana2023}) have shown that most historical OB associations are probably asterisms without kinematic coherence and contaminated by foreground and background members.

Taking advantage of the astrometry from HIPPARCOS, \citet{HipparcosOBAssociations} provided the most comprehensive census of OB associations within 1 kpc at the time, with 13 OB associations divided between 1130 members. A crossmatch between our catalogue and theirs gives 168 stars in common. The breakdown is as follows: we recover 16 (out of 120) of their US members (all of which correspond to our Sco-Cen 2), 42 (out of 221) of their UCC members (divided between our two Sco-Cen associations), 21 (out of 180) of their LCC members (all in our Sco-Cen 1), 20 (out of 93) of their Vela OB2 members (most of which correspond to our Vela OB2), 10 (out of 23) of their Trumpler 10 members (all in our Tr. 10c), 45 (out of 103) of their Collinder 121 members (all in our Collinder 121) and 13 (out of 41) of their Per OB2 members (all in our Per OB2). We thus recover only $\sim$15 \% of their members, which can partly be explained because one third of the members from \citet{HipparcosOBAssociations} are classified as "late-type" (whilst, unsurprisingly, 161 out of the 168 crossmatching members are classified as "early-type"), but mostly because HIPPARCOS astrometry and membership have been superseded by \textit{Gaia}, whose measurements are more precise and which provides access to fainter stars. With the exception of Cas-Tau (which we argue is too extended to be a real OB association, see Appendix \ref{missing}), all the other OB associations from \citet{HipparcosOBAssociations} (Collinder 121, Per OB3, Lac OB1, Cep OB2 and Cep OB6) have been re-discovered with \textit{Gaia} data. In fact, if we crossmatch all the remaining 962 OB association members from \citet{HipparcosOBAssociations} with the 7755 OB stars from \citetalias{Quintana2025} in HIPPARCOS, we find 370 stars in common (356 classified as "early-type"), with members recovered from every OB association in \citet{HipparcosOBAssociations} except Lac OB1, Cep OB2 and Cep OB6. About half of the remaining 760 OB association members from \citet{HipparcosOBAssociations} are classified as "late-type" (compared with one third of the 1130 total number of members), meaning that overall, most of the missing members from \citet{HipparcosOBAssociations} in our catalogue of OB associations can either be attributed to their lack of kinematic coherence once revisited with \textit{Gaia} data (just like for \citealt{MelnikDambis2020}), or because they were too cold to be included in our catalogue.

\citet{BouyAlves2015} built a 3D map of OB stars within 500 pc from the Sun with HIPPARCOS data. In doing so, besides detecting enhanced structures corresponding to well-known OB associations (Sco-Cen, Ori OB1 and Vela OB2), they identified four new OB groups within 500 pc, that they labelled Monorion, Taurion, Orion X, Vela OB5, shared between 123 members. A crossmatch of their members with our OB association members gives only 24 stars in common. Among those, 22 out of the 48 members from Orion X are in Ori OB1b, and we therefore argue Orion X is effectively a subgroup of Ori OB1b, while the others two are in Vela OB5. To contrast, 80 OB members from \citet{BouyAlves2015} are successfully crossmatched with the 7755 OB stars from \citetalias{Quintana2025} in HIPPARCOS (notably including 21 out of the 24 members from Vela OB5), while the majority of the 43 missing stars have been classified as late B-type in \citet{HipparcosOBAssociations}, implying they were not included in \citetalias{Quintana2025} because they were fitted as early A-type stars. We thus confirm that the absence of Taurion, Monorion and Vela OB5 is due to their lack of kinematic coherence once revisited with \textit{Gaia} DR3 data (c.f. Appendix \ref{missing})\footnote{\citet{BouyAlves2015} attibuted the absence of Cas-Tau, Cep OB6 and Lac OB1 to these associations being too loose to be detected by their method. But while Cep OB6 and Lac OB1 have been recovered by some recent studies \citep[e.g.,][]{Kerr2023}, including ours, this has not been the case for Cas-Tau.}.

More recently, \citet{Chemel2022} combined spectroscopic data for O- and early B-type stars (initial mass greater than 20 M$_{\odot}$) from \citet{Skiff2014}, \citet{Xu2018} and the Large Sky Area Multi-Object Fibre Spectroscopic Telescope (LAMOST) DR5 with the young OCs from \citet{Dias2021} to search for OB associations with \textit{Gaia} DR3. In doing so they identified 214 groups split between 5985 stars, that they labeled as `clusters', and correspond to either OB associations or star clusters. A crossmatch between our OB association members and their 1996 members with $\sqrt{X^2+Y^2} <$ 1 kpc gives 431 members in common, as indicated in Table \ref{CompCatalogues}. This weak overlap ($\sim$22 \%) can be attributed to the fact that \citet{Chemel2022} limited their analysis to the most massive stars and relied on spectroscopic data which favours the brightest members of OB associations, introducing some biases in the identification process despite the usage of improved data. We note that out of their 1562 remaining members, 801 (hence more than half of them) successfully crossmatch with our catalogue of SED-fitted OB stars from \citetalias{Quintana2025}. 

\subsection{Star clusters}
\label{starclusters}

Being co-natal and co-moving, OB associations often encompass open clusters (e.g \citealt{Wright2020,Wright2023}). To evaluate the overlap between such stellar groups, the catalogue from \citet{HuntReffert2023} is suited, as it is the largest, most homogeneous and complete catalogue of star clusters to date.  For this work, we exploit the updated version from \citet{HuntReffert2024}, which includes a total of 7167 clusters, including 5647 classified as bound/compact open clusters (hereafter, OCs) and 1520 as unbound clusters/moving groups.

A contrast between the overdensities of our OB stars within 1 kpc and the young, high-quality and compact OCs from \citet{HuntReffert2024} was already performed in \citetalias{Quintana2025}. Here we go further by directly comparing two clustered populations. The crossmatch between our OB association members and the 211,300 star cluster members within 1 kpc from \citet{HuntReffert2024} is displayed in Table \ref{CompCatalogues}, with nearly half our OB association members found in the clusters from \citet{HuntReffert2024}.

The strong overlap between our two catalogues encourages us to exploit these star clusters in order to derive another age estimate for our OB associations, since \citet{HuntReffert2024} calculated isochronal ages for their star clusters. To do so, we have restricted the list of star clusters to the high-quality ones as defined in \citet{HuntReffert2023,HuntReffert2024}, i.e. those with astrometric signal-to-noise ratio larger than 5$\sigma$ and with a median colour-magnitude diagram (CMD) class, $Q_{\rm CMD}$, larger than 0.5. Here, however, we only apply the second condition on clusters older than 20 Myr: as mentioned in \citet{QuintanaHuntParul2025}, differential reddening broadens the CMD of the youngest clusters because many of them are still embedded within their natal gas, therefore it is better to relax the constraint on $Q_{\rm CMD}$ for them. 

Table \ref{TableOC} shows the relation between our 56 OB associations and their related OCs from \citet{HuntReffert2024}, indicating the level of overlap as well as the age of the related OCs and displaying the range of ages spanned by the OCs in case there were several corresponding one. We notice some cases where this range is too large for the OCs to be related together (e.g. Her OB1a, Sco OB3 and Cyg OB10a). Contrariwise, some ages are consistent with the physical age spread observed in OB associations, such as Sco-Cen 1, Vela OB2, Per OB2, Ori OB1b, Sge OB1, Cep OB3 and Vela OB6. Examples of estimated ages with this method aligning well with litterature values are Ori OB1b (4--13 Myr compared with 1--12 Myr in \citealt{Bally2008,Briceno2019,Sanchez2024}),  Cyg OB7 (4--25 Myr compared with $\sim$13 Myr in \citealt{Uyaniker2001}), Cep OB2 (6 Myr compared with 3--16 Myr in \citealt{Szilagyi2023} as well as Her OB1b and Cep OB3 (27 Myr and 4--5 Myr compared with $\sim$29 Myr for Cep-Her and $\sim$3 Myr in \citealt{Kerr2023}). Nevertheless, there are also several examples of OB associations whose ages are over-estimated with this method, such as Per OB3 and Lac OB1 (respectively $\sim$122 Myr and $\sim$28 Myr compared with $\sim$58 Myr and $\sim$10 Myr in \citealt{Kerr2023}).

\subsection{Young stellar groups}

The SPYGLASS survey \citep{Kerr2021} aimed at identifying young ($<$ 50 Myr) stellar groups in the local Milky Way. In \citetalias{Quintana2025}, we  contrasted the overdensities of the clustered population (36,182 stars shared between 116 groups) of its most recent version, SPYGLASS IV \citep{Kerr2023}, with our population of O- and B-type stars.

Here again we proceed as in Section \ref{starclusters} and compare directly our clustered populations. As displayed in Table \ref{CompCatalogues}, we find 313 stars in common. The overlap is smaller than for the star clusters because \citet{Kerr2023} include young stars of all masses, and do include older, late B-type stars whose lifetime can reach hundreds of Myr (e.g. \citealt{Ekstrom}), as evidenced by the median SED-fitted $\log(T_{\rm eff})$ of 4.05 for our 2238 non-matching OB association members (compared with a median value of median SED-fitted $\log(T_{\rm eff})$ of 4.15 for the 313 crossmatching ones).

Among the most common correspondences, we find 91 of our OB association members in their Ori-Per group (split between Ori OB1a, Ori OB1b and Per OB2), 40 in their CaMaS group (corresponding to our Collinder 121), 36 in their Sco-Cen group (corresponding to our Sco-Cen associations), 22 in their Cep-Her group (corresponding to our Her OB1), 22 in their Cep OB2/3 group (corresponding to our Cep OB2a, Cep OB2b and Cep OB3, 18 in their Tr. 10 group (split between Tr. 10a and Tr. 10c), 15 in their Vela group (corresponding to our Vela OB2), 14 in their Cone group (corresponding to our Mon OB1), 12 in their NAS group (corresponding to our Ara OB1) and 10 in their Alessi 20 group (corresponding to our Cas OB15). From this comparison, it is clear that many of our OB associations form larger complexes, with both individual OB associations and young stellar groups from SPYGLASS serving as complementary tracers of the star formation history in the local Milky Way.

\subsection{Star clusters, co-moving groups and structures}
\label{structures}

\citet{Kounkel2020} applied HDBSCAN on \textit{Gaia} DR2 data to identify star clusters, co-moving groups and other structures within 3 kpc, and in doing so detected 8293 stellar groups shared between 987,376 members, and including 1887 groups shared between 351,984 members within $\sqrt{X^2+Y^2} <$ 1 kpc.

We have crossmatched our 2551 OB association members with their 351,984 group members and found 1097 members in common. Just like for Section \ref{starclusters}, this significant overlap constitutes the first step towards a broader analysis of the age estimation of our OB associations, as well as the larger complexes they belong to.

We have applied the same method as in Section \ref{starclusters} to analyse the overlap between our OB associations and the stellar groups from \citet{Kounkel2020}, and the results of this process are displayed in Table \ref{TableGroups}. Just like Table \ref{TableOC}, some of the age ranges can reflect real physical spread in ages observed in OB associations (e.g. Sco-Cen and Per OB2), whereas others rather indicate the corresponding stellar groups are likely unrelated (e.g. Cas OB15 and Cam OB1).

Table \ref{TableGroups} provides some complementary information with respect to Table \ref{TableOC}, particularly as \citet{Kounkel2020} compared their identified groups with \textit{Gaia}-based catalogues of OCs that were available at the time \citep[e.g.,][]{CantatGaudin2018,CantatGaudin2019c,CastroGinard2019,CastroGinard2020}. Some unsurprising connections appear, such as the Sco-Cen associations with Corona Australis and Upper Sco or Ori OB1 with the larger Orion complex, whilst others OB associations match directly with the corresponding one in \citet{Kounkel2020} (Cep OB6, Per OB2, Vela OB2, Trumpler 10 and Cep OB3). More surprising, our Per OB3 appears to be related to the Pleiades rather than $\alpha$ Per (which is related to Cep OB6 instead), although an overdensity of B-type stars in the Pleaides was already visible in previous maps \citep[e.g.,][]{BouyAlves2015}. 

Just like Table \ref{TableOC}, the estimated ages from the related groups from \citet{Kounkel2020} should be considered as upper limits instead of accurate estimations. While some of these ages are close to past literature estimates, such as Sco-Cen (7--13 Myr compared with 3--19 Myr in \citealt{MiretRoig2022,Ratzenbock,Posch2025}), Ori OB1b ($\sim$6 Myr compared with 1--12 Myr in \citealt{Bally2008,Briceno2019,Sanchez2024}) and Per OB2 (4--8 Myr compared with 2--7 Myr in \citealt{Bally2008,Kounkel2022}), many OB associations have wide age intervals, some of which older than 100 Myr even though OB associations are thought to dissolve after a few tens of Myrs \citep{Wright2020}.

\section{Galactic structure}
\label{structure}

OB associations are important for our understanding of the star formation process \citep[e.g.,][]{Wright2023}, and thereby can be served to trace the position and motion of the spiral arms \citep[e.g.,][]{Quintana2023}. Nevertheless, our current census is limited to the nearest 1 kpc, closer than the known locations of the Perseus and Carina-Sagittarius arms \citep[e.g.,][]{Reid2019}. Recent studies have confirmed the potential of using overdensities of O- and B-type stars to trace the Galactic spiral arms, including the Local one \citep[e.g.,][]{Ge2024}, but they need to be complemented by other tracers of star formation in the solar neighborhood, which can themselves be connected to larger scales.

Simulations suggest the existence of wave-like corrugations in disc galaxies \citep[e.g.,][]{Widrow2012,Carlin2013,Williams2013,TepperGarcia2022}. In the Milky Way, they have been linked to the passage of the Sagittarius dwarf galaxy \citep[e.g.,][]{Purcell2011,Gomez2013,Laporte2018,Antoja2018,Laporte2019,BlandHawthornTepperGarcia2021}. Using Auriga cosmological simulations, \citet{Gomez2017} showed that these corrugations get imprinted both in the gas and in the stars, including young stellar populations, which was later corroborated by \citet{TepperGarcia2022}.

It is therefore not surprising that such vertical corrugations have been related to the Radcliffe Wave and its stellar counterpart such as the Cepheus Spur \citep{PantaleoniGonzalez2021}, and confirmed with the discovery of a distant corrugation in the Milky Way disk referred to as `The great wave' \citep{Poggio2025}. In this context, structures similar to the Radcliffe Wave serve as tracers of star formation in the local Milky Way.

\citet{Kormann2026} exploited the 3D extinction map from \citet{Edenhofer2024} to identify elongated, parallel structures in the local Milky Way referred to as superclouds, a nomenclature originally proposed by \citet{ElmegreenElemegreen1983} to designate the regular distribution of star-forming complexes and their associated neutral atomic hydrogen (H$_{\rm I}$) in external galaxies, and subsequently observed in the inner Milky Way \citep{Dame1986,ElmegreenElmegreen1987}. In addition to the previously-known Radcliffe Wave \citep{Alves2020} and Split \citep{Lallement2019}, \citet{Kormann2026} discovered five new superclouds, labeled as the Sagittarius Spur Extension (SSE), Malpolon Cloud, Vela Ridge Cloud, Natrix Cloud and Anguis Cloud, with all but the last one located within $\sqrt{X^2+Y^2} <$ 1 kpc. 

\begin{figure*}
    \centering
    \includegraphics[scale = 0.52]{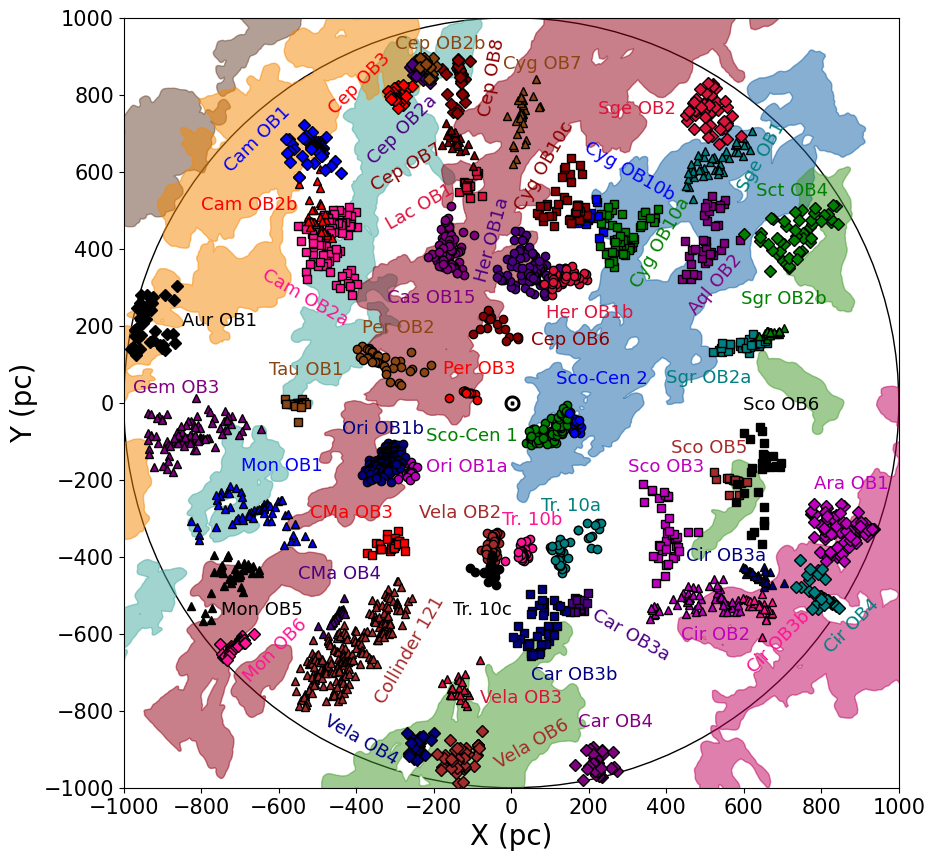}
    \caption{Same as the top panel from Fig. \ref{Map_OBAssociations} but with the superclouds from \citet{Kormann2026}, which includes the Radcliffe Wave (in red, \citealt{Alves2020}) and the Split (in blue, \citealt{Lallement2019}). The other newly identified superclouds on this figure are the Malpolon Cloud (in orange), Natrix Cloud (in cyan), Vela Ridge Cloud (in green) and SSE (in pink).}
    \label{Map_Superclouds}
\end{figure*}

\begin{figure*}
    \centering
    \begin{minipage}{.45\textwidth}
    \includegraphics[width=\textwidth]{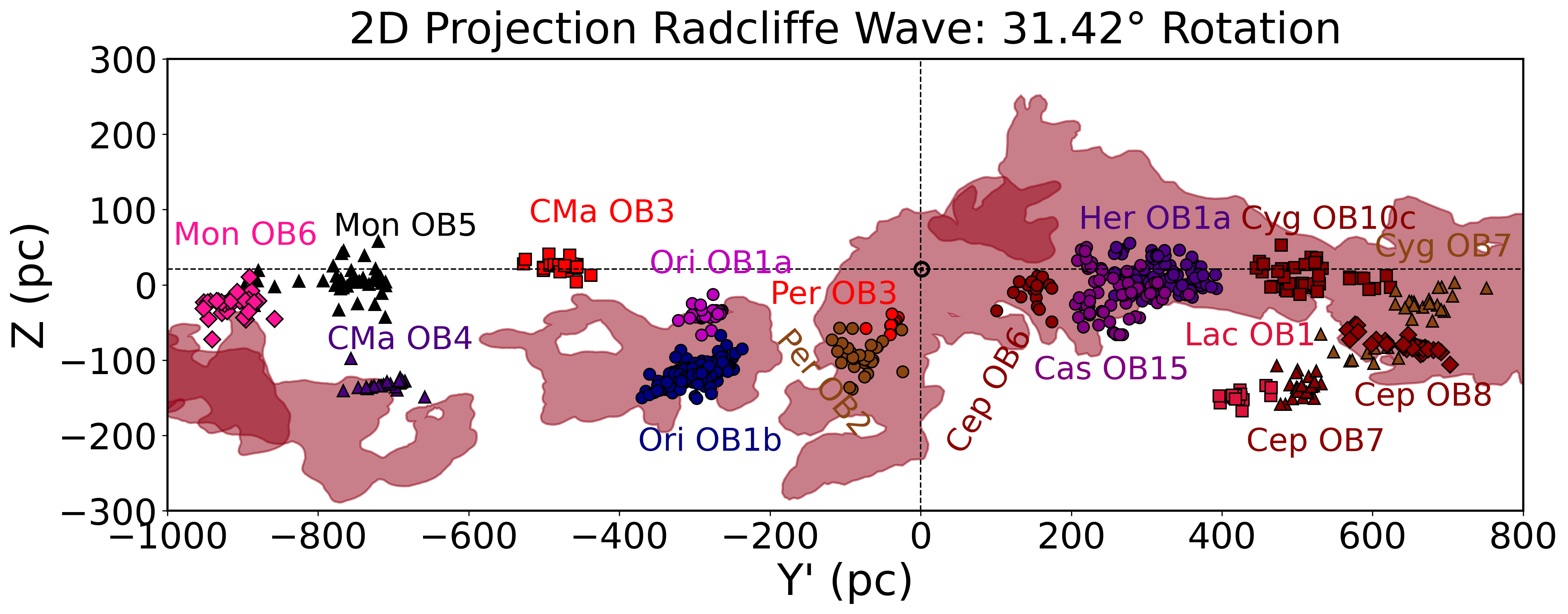}
    \includegraphics[width=\textwidth]{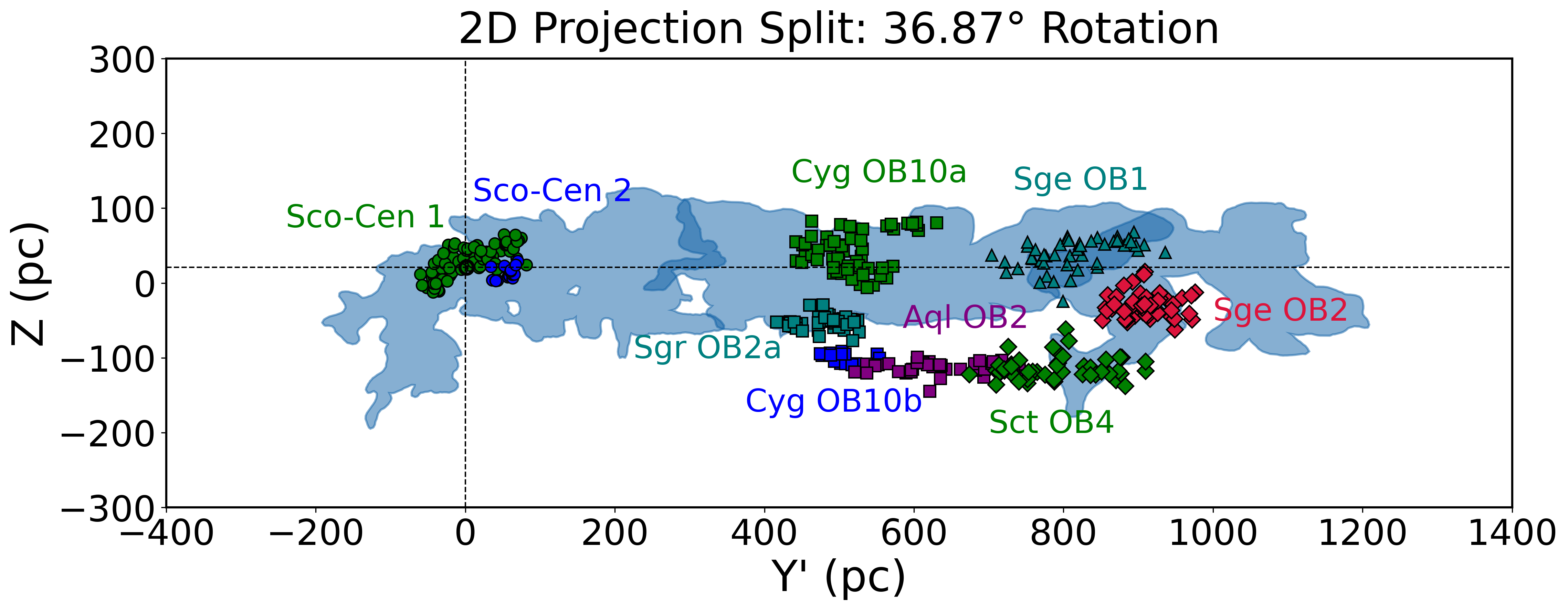}
    \includegraphics[width=\textwidth]{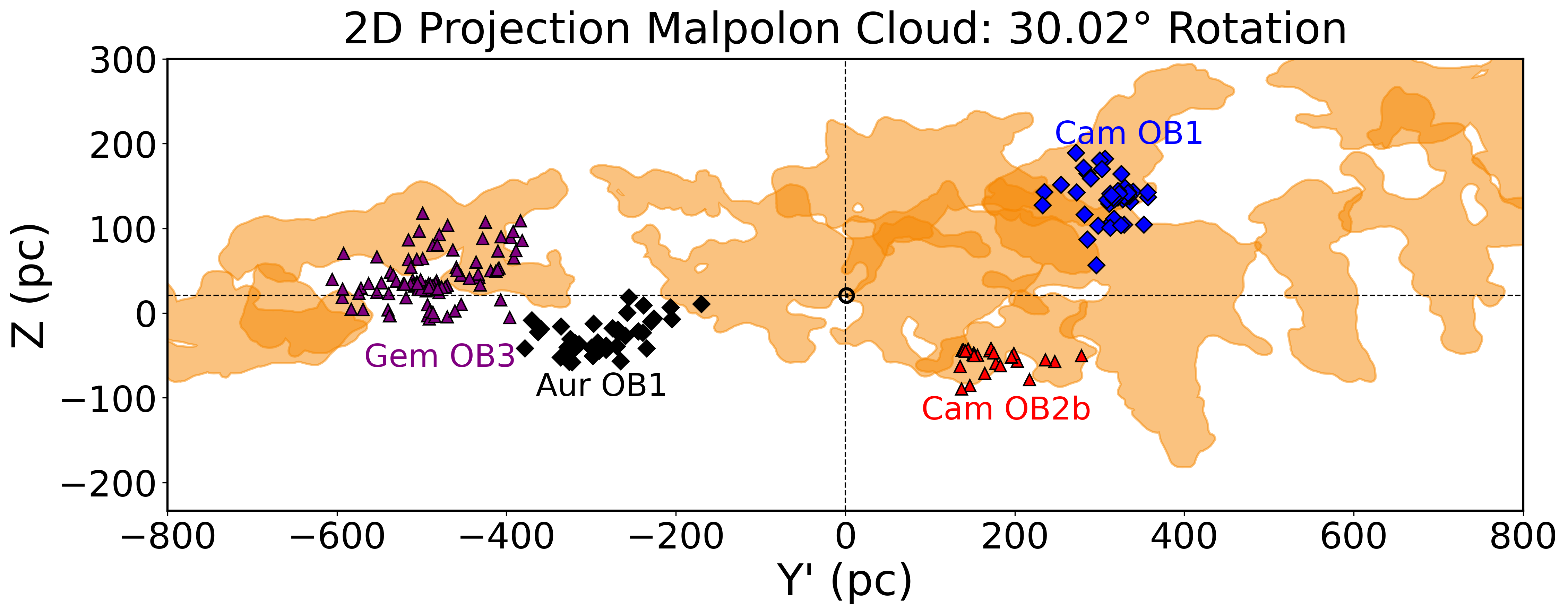}
    \end{minipage} 
     \begin{minipage}{.45\textwidth}
    \includegraphics[width=\textwidth]{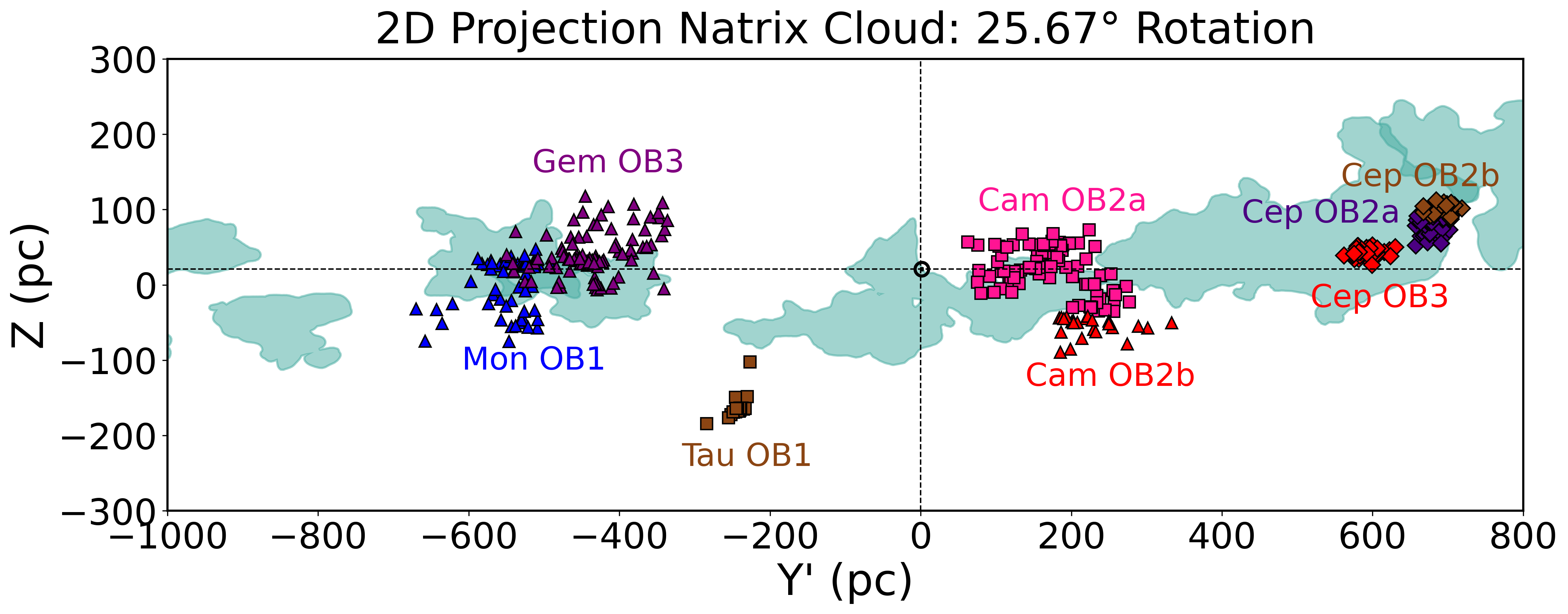}
    \includegraphics[width=\textwidth]{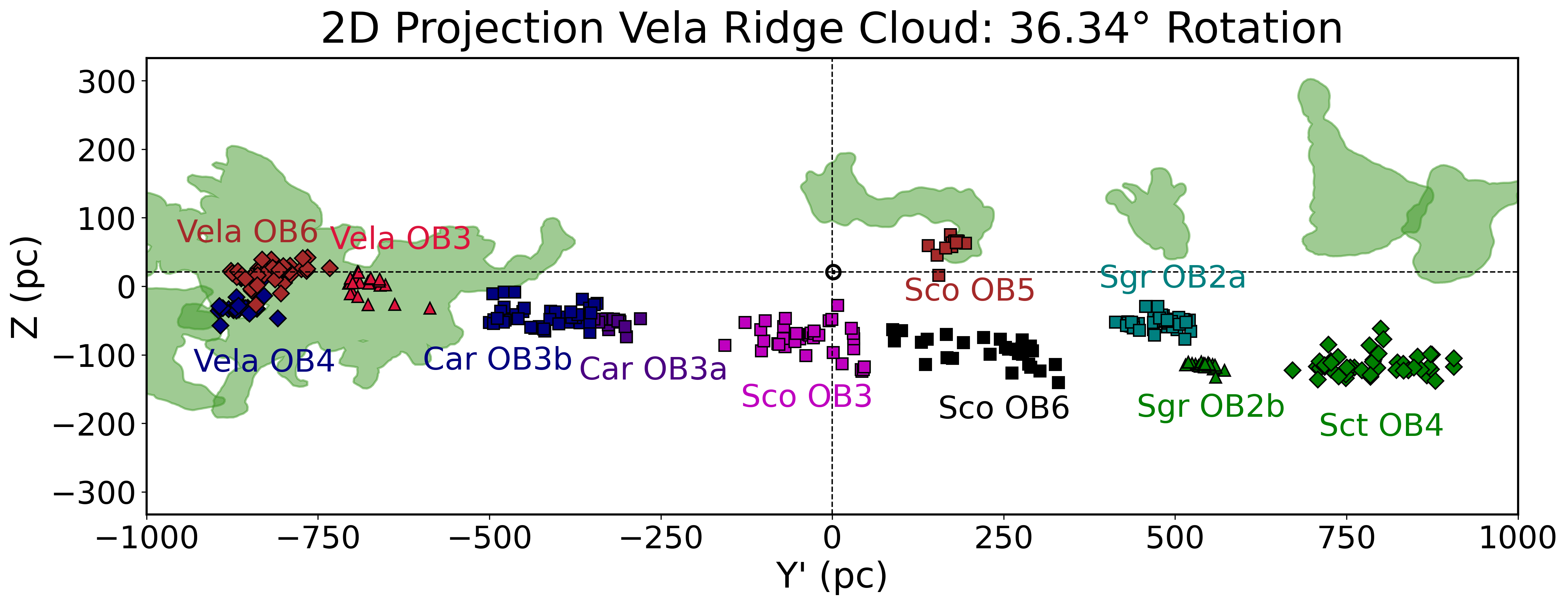}
    \includegraphics[width=\textwidth]{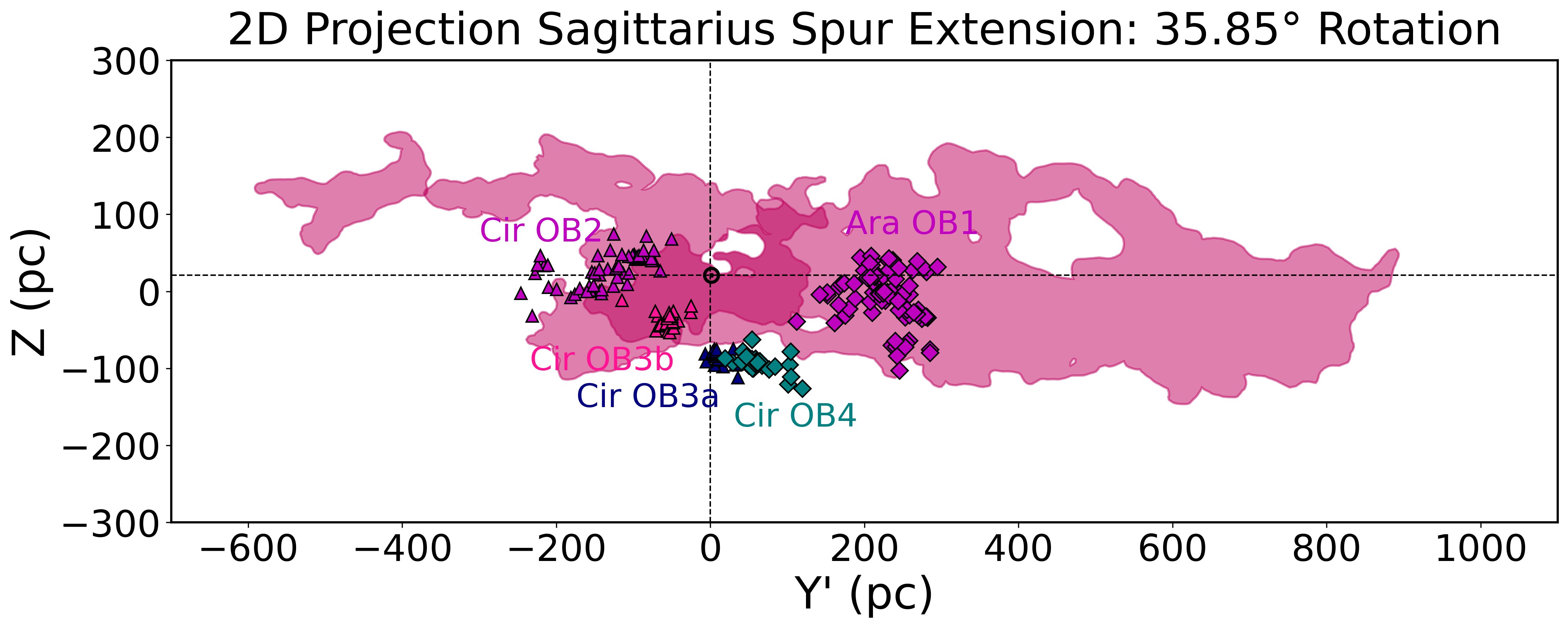}
     \end{minipage} 
    \caption{2D projection on the Z/Y' plane of the six superclouds from \citet{Kormann2026} within 1 kpc, rotated clockwise, with the OB associations displayed on top and rotated with the same angle. The location of the Sun is indicated with the corresponding symbol, using $Z_{\odot} = 20.8 \pm 0.3$ pc from \citet{BennettBovy2019}.}
    \label{YZ_Superclouds}
\end{figure*}

This dataset, combined with our new census of OB associations, enables us to provide a first overview of the correlation between the superclouds and young stellar tracers. We have thus plotted the location of our OB associations on top of the superclouds from \citet{Kormann2026} in the X-Y plane in Fig. \ref{Map_Superclouds}. This picture reveals that both the superclouds and the OB associations are ubiquitous across this face-on view of the local Milky Way, preventing us from establishing direct connections. However, there are also several noticeable cases worthy of comment:
\begin{itemize}
    \item \textbf{Vela OB2 and Trumpler 10}: Both OB associations form the largest sequence that are not related to any supercloud (also evident in Fig. \ref{YZ_Superclouds}). Given their older ages (20--50 Myr), it is unsurprising that they are now decoupled from their natal gas (with the exceptions of Vela OB2 that includes the WR star $\gamma^2$ Vel, as discussed in Section \ref{maxages}).
    \item \textbf{Her OB1}: These two OB associations are part of the bigger Cep-Her, originally identified in \citet{Prisinzano2022}. \citet{Kerr2023} estimated an age of $\sim$29 Myr for the complex, effectively acting as a bridge between the Radcliffe Wave and the Split. Broken down to individual components, Her OB1a is associated with the Radcliffe Wave, but Her OB1b fits this bridging role.
    \item \textbf{Gem OB3:} is one of our most extended OB associations, both in positional and velocity space. Its hottest star, HD 249388, was fitted with $\log(T_{\rm eff}) = 4.15^{+0.04}_{-0.03}$, corresponding to a mid B-type star \citep{Mamajek}. As Gem OB3 lies between the Malpolon Cloud and the Natrix Cloud, this is consistent with the picture of an OB association whose massive stars have already died out, blowing away their surrounding gas through feedback.
    \item \textbf{Sct OB4:} likewise, this OB association lies within the ring that separates the Split from the Vela Ridge Cloud. Its hottest star, HD 182975, was fitted with $\log(T_{\rm eff}) = 4.24^{+0.08}_{-0.04}$, which corresponds to a B2-B3 star \citep{Mamajek}. Unveiled is the picture of an OB association of moderate age, whose feedback is currently clearing away its neighboring gas, as its furthest members from its centre are located at the edges of the superclouds.
    
\end{itemize}

In the next subsection, we will focus on individual superclouds to further study these correlations in positional space, particularly with the undulations that the Radcliffe Wave, Malpolon Cloud, Natrix Cloud and the Vela Ridge Cloud form in the Y-Z planes, as displayed in Fig. \ref{YZ_Superclouds}. We however stress that we are comparing their current 3D position: the kinematics of the superclouds will be studied in a future work (Kormann et al., in prep.): combined with a more accurate determination of isochronal and kinematic ages of OB associations, this will allow us to determine the mechanisms in play (i.e., whether OB associations decoupled from the natal gas moved away from them or blew away their surrounding gas). Only a full kinematical analysis, akin to the approach from \citet{Konietzka2024} for the Radcliffe Wave, would be able to confirm the physical connections between some OB associations and the superclouds.

\subsection{The Radcliffe Wave}
\label{radcliffe}

The Radcliffe Wave is a 3D structure initially identified in \citet{Alves2020} from the molecular cloud census of \citet{Zucker2020}. It is shaped as a coherent alignment of dense gas that extends from Canis Major to Cygnus-X and intercepts several major star-forming regions such as Orion. Since its discovery, the Radcliffe Wave has been traced by young stars \citep{Tu2022,Kerr2023} and OCs \citep{Konietzka2024,Bobylev2025}, it is therefore natural to explore whether our OB associations act as tracers of the Radcliffe Wave as well. 

In Fig. \ref{YZ_Superclouds}, we have plotted the OB associations from Fig. \ref{Map_Superclouds} that are close to the Radcliffe Wave across the X-Y plane, here projected on the Z/Y' plane. This picture highlights the necessity of the 3D view as some OB associations close to the Radcliffe Wave in Fig. \ref{Map_OBAssociations} are actually above or below it, such as Mon OB5, Mon OB6, Cep OB7 and Lac OB1, which, in the latter case, is consistent with the findings of \citet{Konietzka2024}. Some spatial correlations between our OB associations and the Radcliffe Wave are corrborated here: CMa OB4,  Ori OB1a, Ori OB1b, Per OB2, Per OB3, Cep OB6, Her OB1a, Cyg OB10c, Cep OB8 and Cyg OB7. Some of these correlations were expected as the Canis Major, Orion, Perseus and Cepheus massive star-forming complexes are known too be part of the Radcliffe Wave. This is consistent with OB associations whose different age estimators from Section \ref{analysis} and Section \ref{comp} confirm their youth, such as Ori OB1b and Cyg OB7, but other OB associations will require more accurate age determinations and kinematics to fully connect them to the Radcliffe Wave.

\subsection{The Split}
\label{split}

The Split is an inter-arm feature, originally identified in \citet{Lallement2019} as a spur longer than 2 kpc spanning from the Local to the Carina-Sagittarius spiral arm, standing out in their 3D dust map. In their study, they reported an overdensity of molecular clouds along the Split, but failed to notice a correlation with  H{\sc ii} regions, masers and O-type stars. As such, our census of OB associations represents an opportunity to analyse again the connection between this spur and young stellar populations, especially as more recent papers identified Sco-Cen as part of the Split \citep[e.g.,][]{MiretRoig2022}.

We have also displayed the location of the Split on top of our OB associations in the X-Y plane in Fig. \ref{Map_Superclouds}. Compared with historical censuses of OB associations \citep{Wright2020}, we have identified for the first time in this work (to the best of our knowledge) OB associations on the first quadrant that could serve as potential tracers of the Split further than Sco-Cen, such as Aql OB2, Cyg OB10a, Sge OB1 and, to a lesser extent, Sge OB2. This was hinted in \citet{Swiggum2022}, where the Split was described as being formed from the Sco-Cen, Aquila and Serpens star-forming clouds.

However, Fig. \ref{YZ_Superclouds} shows that Aql OB2 is actually located too far from the Galactic Plane to be connected to the Split, contrary to Sge OB1, Cyg OB10a and Sge OB2. But again, our lack of accurate age estimations alongside precise 3D kinematics for these new OB associations prevent us from confirming this connection, warranting a future full kinematic investigation.

\subsection{The other superclouds}
\label{othersuperclouds}

\begin{itemize}
    \item \textbf{Malpolon Cloud:} Only Cam OB1 (and Cam OB2b, to a lesser extent) shares its 3D position with this supercloud, although Gem OB3 acts as bridges between Malpolon and Natrix.
    \item \textbf{Natrix Cloud:} The Cepheus OB associations are fully located into this cloud, which would be consistent with their young ages ($<$ 15 Myr) estimated in this study. While we could not determine as accurately the age of Mon OB1, it is also known as a very young association ($<$ 10 Myr, e.g. \citealt{Lim2022}) and contains many members in the Natrix Cloud. As for Gem OB3, \citet{Kormann2026} mentioned that the feedback from high-mass stars could shear off individual clouds  \citep{Zari2023,PantaleoniGonzalez2025}: consequently, and as it bridges the Natrix Cloud and Malpolon Cloud, it could be an older association which has played this role in the past.
    \item \textbf{Vela Ridge Cloud:} The three Vela OB associations (Vela OB3, Vela OB4 and Vela OB6) occupy the centre of the largest cloud. However, only Vela OB6 appears as very young from all the different age estimation methods of this paper, while this region is a well-known complex associated to the supershell GSH 238+00+09 \citep{Heiles1998}. Elsewhere, only the 3D position of Sco OB5 coincides with the supercloud, but \citet{Kormann2026} mentioned that the Vela Ridge Cloud could be an extension of the SSE rather than an independent supercloud, and its components on the upper-right part of Fig. \ref{Map_Superclouds} a part of the Split instead.
    \item \textbf{SSE:} Fig. \ref{YZ_Superclouds} confirms that Cir OB2, Cir OB3b and Ara OB1 share their 3D position with this superclouds, contrary to Cir OB3a and Cir OB4 which are located at their edges. Cir OB2 and Cir OB3a have kinematic ages consistent with being $<$ 20 Myr whilst Cir OB2 and Ara OB1 have maximum isochronal ages younger than 30 Myr, but the disagreement between their kinematic and maximum isochronal ages suggest that a more thorough kinematical analysis required to confirm their connection with the supercloud. Still, since \citet{Kormann2026} identified this supercloud as an extension of the spur of the Carina-Sagittarius arm discovered by \citet{Kuhn2021}, which is itself traced by OB stars \citep{PantaleoniGonzalez2025}, it is expected that the SSE is also traced by OB stars, as it encompasses the Circinus complex \citep{Kerr2023,Kerr2025}.
\end{itemize}

\section{Conclusions}
\label{conclusions}

In this paper we have exploited a complete census of $\sim$25,000 O- and B-type stars within 1 kpc, to which we have applied a clustering algorithm to their 3D positions and 2D velocities. In doing so we have identified 56 highly-confident OB associations, doubling the census in this volume compared with the known number of 28 OB associations within 1 kpc from \citet{Wright2020}.

We have characterized these OB associations physically and kinematically, and estimated linear velocity gradients compatible with the majority of them ($\sim$68 \%) expanding in at least one direction. We contrasted our OB association members with other catalogues of OB associations, young stellar groups and star clusters, as well as with the superclouds of the local Milky Way, in order to trace star formation in the solar neighborhood. Because our catalogue of OB associations match well with existing features, we argue that it is highly-reliable and supersedes previously-made general catalogues of OB associations within 1 kpc.

The present study was focused on the current location of the OB associations. Because we have only identified their upper-mass members, the lack of reliable RVs prevented us from analysing their full 3D kinematics and to estimate reliable isochronal ages. The upcoming releases of the William Herschel Telescope Enhanced Area Velocity Explorer (WEAVE, \citealt{WEAVE}) and the 4-metre Multi-Object Spectroscopic Telescope (4MOST, \citealt{4MOST}), will allow us to obtain more reliable RVs for the hottest members of these OB associations. Complementary to this, we intend to extend the membership to lower-mass stars, which will enable a more precise estimation of their isochronal and kinematic ages. In the future we intend to perform a full 3D orbital analysis of the OB associations to trace their place of birth and to fully link them to surrounding features, unveiling a full picture of the star formation history of the local Milky Way. An accurate catalogue of OB associations with reliable isochronal and kinematic ages will be statistically-significant enough to study the timescale between the formation of stars and the assembly of stellar groups \citep[e.g.][]{MiretRoig2024}.

As we are also extending the census of OB stars up to a distance of 2 kpc (Quintana et al., in prep.), we aim at expanding the census of OB associations as well, using them as tracers of the position and motions of the Galactic spiral arms.

\section*{Acknowledgements}

ALQ acknowledges a PSL fellowship granted by the Scientific Council from the Paris Observatory. This work was supported by CNES, focused on the Gaia mission. CFPL acknowledges funding from the European Research Council (ERC) under the European Union’s Horizon 2020 research and innovation programme (grant agreement No. 852839) and the Agence Nationale de la Recherche (ANR project ANR-24-CPJ1-0160-01).

The authors would like to thank the anonymous referees for their insightful comments that helped us to improve the paper. The authors also would like to thank Sergio Sánchez-Sanjuán for providing his catalogue of Ori OB1 members, Josefa Großschedl for her help to best display the 3D extinction map, Emily Hunt for explanations on how to best exploit her catalogue of open clusters, Juan Martínez García for discussions on the kinematics of OB stars, Joss Bland-Hawthorn for his insights on Galactic dynamics that helped improving the scientific discussions of the paper, Marc Audard for his recommendations of external catalogues to compare the catalogue of OB associations with, Hervé Bouy for clarifications regarding his catalogue of OB groups, Kevin Jardine for insightful discussions about the nomenclature of OB associations, and Joseph Armstrong for insightful discussions about the kinematics of stellar groups.

This work has made use of data from the European Space Agency (ESA) mission
{\it Gaia} (\url{https://www.cosmos.esa.int/gaia}), processed by the {\it Gaia}
Data Processing and Analysis Consortium (DPAC,
\url{https://www.cosmos.esa.int/web/gaia/dpac/consortium}). Funding for the DPAC
has been provided by national institutions, in particular the institutions
participating in the {\it Gaia} Multilateral Agreement.

This work also benefited from the use of \textit{TOPCAT} \citep{Topcat}, Astropy \citep{Astropy} and the Vizier and SIMBAD database, both operated at CDS, Strasbourg, France.

\section*{Data Availability}

The full list of the 56 OB associations and their properties (as described in Table \ref{Table_OBAssociations}) will be uploaded to Vizier, as well as their individual members, and the groups initially identified in Section \ref{identification} that were ultimately rejected.



\bibliographystyle{mnras}
\bibliography{bibliography} 




\appendix

\section{Choice of parameters for the clustering algorithm}
\label{parameter_space}

As described in Section \ref{clustering}, the HDBSCAN outputs depend critically on the adopted values of its inputs: in this section, we detail how we explored the parameter space to find the best configuration that leads to an identification of stellar groups physically consistent with being OB associations.

Firstly, the parameter \texttt{cluster\_selection\_method} could either be `excess of mass (EOM)' or `leaf'. Here we favour the second method that finds smaller and more homogeneous groups, and has therefore been consistently used to search for OB associations (e.g. \citealt{SantosSilva2021,Chemel2022,Quintana2023}).

Secondly, \texttt{min\_samples} is the parameter used to compute the distance between a point and its nearest neighbors \citep{HDBSCAN}. In a past work \citep{Quintana2023}, we adopted \texttt{min\_samples}  = 10, as this resulted in groups with a stellar density consistent with OB associations, we thus use the same value here as our overall density of O- and B-type stars selected from our SED fitter in \citetalias{Quintana2025} will be similar.

The choice for the other parameters was less straightforward. \texttt{min\_cluster\_size} corresponds to the minimum number of OB stars (in our case) within an OB association. In \citet{Quintana2023}, we chose a value of 15, following the typical number of OB stars found in historical OB associations \citep{Humphreys1978}, and below which there is an increasing risk of noise. However, systems with fewer OB stars could be T associations (whose name originates from the T-Tauri stars that are their most prominent members). For this study, we follow the definition from \citet{Wright2020} and consider OB associations and T associations to be distinct entities, thereby adopting a minimum value of 15 OB stars for \texttt{min\_cluster\_size}. \texttt{cluster\_selection\_epsilon}, on the other hand, is defined as a distance threshold \citep{MalzerClaudia2019}, which allows small groups with similar properties to be connected and thus merged (as in e.g. \citealt{Kerr2021,Kerr2023}).

We have then have varied \texttt{min\_cluster\_size} between 15 and 30 (with a step size of 5), \texttt{cluster\_selection\_epsilon} between 20 and 40 (with a step size of 10) and the normalization factor $c$ between 4 and 6 pc km$^{-1}$ s$^{-1}$. This choice of parameter spaces ensured to obtain groups physically corresponding to OB associations, and led us to a total of 36 different configurations to explore.

On the one hand, the effects of increasing \texttt{min\_cluster\_size} was straightforward, as it resulted in the disappearance of smaller groups, some of which corresponded to well-known OB associations such as Sco-Cen, Per OB3 and Cep OB6. We thus rapidly concluded that a lower value of \texttt{min\_cluster\_size} increased our completeness whilst still allowing us to identify larger OB associations such as Ori OB1, a similar trend as in \citet{Quintana2023}. Consequently, we have adopted a value of \texttt{min\_cluster\_size} of 15.

Likewise, whilst varying \texttt{cluster\_selection\_epsilon}, we quickly noticed that it led to the merging of smaller groups and resulted in groups too extended in positional space to correspond to OB associations, encouraging us to adopt a smaller value for this parameter.

Nevertheless, varying the normalization factor $c$ led to more subtle effects, so to find the best configuration, we carried out a MC simulation where we randomly sampled the Galactic Cartesian coordinates and the transverse velocities of the individual members of the groups within their uncertainties. At each iteration, we recorded the dispersion (defined as one standard deviation) in each of their 5D parameters ($X$, $Y$, $Z$, $V_l$, $V_b$), after which we subtracted this observed dispersion by the median uncertainties of the group to obtain an intrinsic dispersion. We repeated this process 1000 times to obtain a median dispersion for each parameter and for each group\footnote{Our method to estimate the intrinsic velocity dispersions in OB associations therefore neglects binary motion, but its effects on the transverse velocities are much smaller than for the radial velocities. \citet{Jackson2020} estimated that neglecting binary motions for the transverse velocities would only lead to a $\sim$1 \% under-estimation of the number of cluster members, as illustrated in their Fig. 7.}.

OB associations are kinematically-coherent stellar groups (no more than a few km s$^{-1}$ in velocity dispersion) and that typically extend over a few tens of pc \citep{Wright2020,Wright2023}. These properties governed our choices of HDBSCAN inputs. The low density of OB stars prevents us from identifying groups compact enough to identify star clusters, therefore the risk is to identify too many groups that are too extended (in positional or velocity space) to be OB associations.

To verify this, for each configuration, we have calculated the percentage of identified groups with an intrinsic positional dispersion above 50 pc or an intrinsic velocity dispersion above 4 km s${-1}$ in at least one dimension (the conservative false positive rate), as well as the percentage of groups with a positional dispersion above 100 pc or an intrinsic velocity dispersion above 5 km s${-1}$ in at least one dimension (the liberal  false positive rate). We ultimately chose the configuration that gave us both the smallest liberal and conservative false positive rates, corresponding to \texttt{min\_cluster\_size} = 15, \texttt{cluster\_selection\_epsilon} = 30 and $c$ = 6 km s$^{-1}$).

\section{Contaminants and incompleteness in the clustering process}
\label{contaminants_incompleteness}

In this section we detail how we discarded the kinematical false positive found among the original 102 candidate OB associations identified in Section \ref{clustering} (Section \ref{filtering_kinematical}). Given that this method can be too restrictive, particularly for nearby groups whose membership partly relies on HIPPARCOS astrometry, we have also outlined how we added back some false negative groups (Section \ref{false_negatives}).

\subsection{Quantifying the reliability of each group}
\label{filtering_kinematical}

We estimate the probability of each group following the method of \citet{Quintana2023}. We perform an MC simulation and at each iteration of the MC simulation, we randomize the 5D parameters within their uncertainties and applied HDBSCAN again toall stars in our sample. We consider the resulting groups (which can therefore have a different membership and number of stars compared with the original groups) to match the original groups if all their median 5D parameters are located within 1$\sigma$ of the median 5D parameters of the original group. In doing so we also record the frequency with which each of the individual group members are found in each iteration of the MC simulation. 

We repeated this bootstrap process over 1000 iterations: the probability of a group is thus defined as the number of times an original group is recovered, divided by the number of iterations, and the membership probability of an individual star is calculated as the fraction of times it appeared in a matching group.

To determine which probability threshold to use to identify highly-reliable groups, we also proceeded as in \citet{Quintana2023}, to quantify the probability of false positive groups being detected. We performed another MC simulation by creating a randomized field of 24,706 OB stars, whose 5D parameters have been generated as follows. $X$ and $Y$ were drawn from a randomized uniform distribution within a circle of 1 kpc radius, whereas $Z$ was drawn from a random normal distribution with the median and standard deviation equal to the observed values, and $V_b$ uniformly randomized between -1$\sigma$ and 1$\sigma$ from the observed distribution in order to avoid the effects of outliers.

The distribution of $V_l$ for these synthetic stars, however, requires a more accurate model, given that stellar motion in the Galactic longitude position depends on direction because of the effect of Galactic rotation \citep[e.g.,][]{Almannaei2024}. In such a framework, $V_l$ is provided by the following equation:

\begin{equation}
V_l = V_\ell^{gal} -V_{R,\odot}\sin(\ell)-V_{\phi,\odot}\cos(\ell).
\end{equation}
\noindent where $V_\ell^{gal}$ stands for the Galactocentric transverse velocity in the $l$ direction, and the constants $V_{R,\odot}$ and $V_{\phi,\odot}$ are defined as the velocity of the Sun moving away from the Galactic centre and in the azimuthal direction, respectively. For these quantities, we have adopted the values of $V_{R,\odot}=-10\pm1$ km s$^{-1}$ from \citet{Bland2016} and $V_{\phi,\odot}=247.4\pm 1.4$ km s$^{-1}$ from \citet{GRAVITY2019}.

$V_\ell^{gal}$ is well approximated with the following model of Galactic rotation:

\begin{multline}
    V_\ell^{gal}\simeq \left(V_{\phi,0} + \dfrac{dV_\phi}{dR}\Big{|}_0(R-R_{gal,\odot})\right) \cos(\phi+\ell) \hspace{0.5cm} + \\ \left(V_{R,0}+ \dfrac{dV_R}{dR}\Big{|}_0(R-R_{gal,\odot})\right)\sin(\phi+\ell) \equiv V_\ell^{model}(R,\ell)
    \label{eq:model};
\end{multline}

\noindent where we have used the cylindrical coordinates $R = \sqrt{(R_{gal,\odot}-X)^2+Y^2}$ and $\phi=\arctan{\dfrac{Y}{R_{gal,\odot}-X}}$, adopting the value of  $R_{gal,\odot}=8178\pm 25$ pc from \citet{GRAVITY2019} for the distance of the Sun to the Galactic Centre. $R$ and $\phi$ are derived from the randomized values of $X$ and $Y$ just like $l$. Finally, we have taken the values from the fit of the OB stars within 1 kpc in \citetalias{Quintana2025}, displayed in Table 1 from Mart\'inez Garc\'ia et al. (submitted).

This process gave us a list of synthetic OB stars with randomized XYZ and V$_{l,b}$, to which we added synthetic noise. To that end, we have generated a randomized uniform distribution of uncertainties in equatorial coordinates, PMs and distances between 0 and 3$\sigma$ from their observed distribution, which were then propagated. We have performed this synthetic bootstrap over 1000 times for each set of associations, recording the probability and median distance for each.

These simulations have produced a total of 67,785 synthetic groups ($\sim$68 per simulation, as we carried out 1000 simulations), the overwhelming majority of which have small probability, and only 269 of them have a probability over 50\%, most of which are closer than 500 pc. This thus leads to a false positive rate of $\sim$0.5\%, so we can safely choose a probability of 50\% as the threshold for our observed groups, reducing our list to 62 candidate OB associations.

\subsection{Estimating our incompleteness}
\label{false_negatives}

The method detailed in Section \ref{filtering_kinematical} produced low probabilities for some groups that are clearly matched to some existing and well-studied nearby OB associations (e.g., Sco-Cen and Per OB3). These associations contain a significant number of OB members that lack \textit{Gaia} astrometry, whose only valid astrometry is from HIPPARCOS \citep{vanleuwen2007}, which is of a lower quality than \textit{Gaia} DR3 (see Section 2.1.2 from \citetalias{Quintana2025}): in an OB association at a distance of $\sim$100 pc, the uncertainties on positions are typically 3--4 times larger in HIPPARCOS than in \textit{Gaia} DR3, while those on velocities are typically 2--2.5 times larger. 

To determine the completeness of our OB association census, and effectively to test whether our methods can reliably recover the nearest OB associations, we conducted the following simulation. We generated the members of an OB association with the same 3D positions (XYZ) and velocities (UVW) as Lower Centaurus Cruw (LCC), using 1$\sigma$ values from \citet{WrightMamajek2018}. 

The synthetic noise on these parameters was sampled from the observed errors of the group in our sample that matches Sco-Cen, and scaled with distance, assuming a spherical OB association (i.e. the synthetic uncertainties at d = 1000 pc are five times larger than at d = 100 pc for $X$,$Y$,$Z$,$V_l$ and $V_b$). For simulated groups closer than 300 pc, we generated an IMF from \citet{IMFMasch}, from which we obtained a set of apparent magnitudes $G$, using the stellar evolutionary models from \citet{Ekstrom} and the \textit{Gaia} DR3 PARSEC isochrones \citep{Chen2015}, assuming a unique central distance $d$ and neglecting extinction. We assigned HIPPARCOS uncertainties for members with $G < 4.5$ mag (the brightest magnitude for a star in the Sco-Cen 1 group with valid \textit{Gaia} data), and \textit{Gaia} uncertainties otherwise.

The members of this test association were then placed in a randomized field of OB stars generated as described in Section \ref{filtering_extension}. For each iteration of this simulation, we randomized a new field of OB stars, as well as the positions and velocities of the OB stars of the test association within their uncertainties. We then applied HDBSCAN and verified whether any of the identified groups corresponded to the test association, i.e. whether it was properly recovered. We repeated this process for 100 iterations, varying the number of members from N = 15 (the minimum number of OB members of an OB association, see Section \ref{clustering}) to 100 with a step size of 5 between 15 and 50 and a step size of 10 between 50 and 100, and the median distance from 25 to 1000 pc with a step size of 50 pc between 25 and 300 pc, and a step size of 100 pc between 300 and 1000 pc.

\begin{figure}
    \centering
    \includegraphics[scale = 0.3]{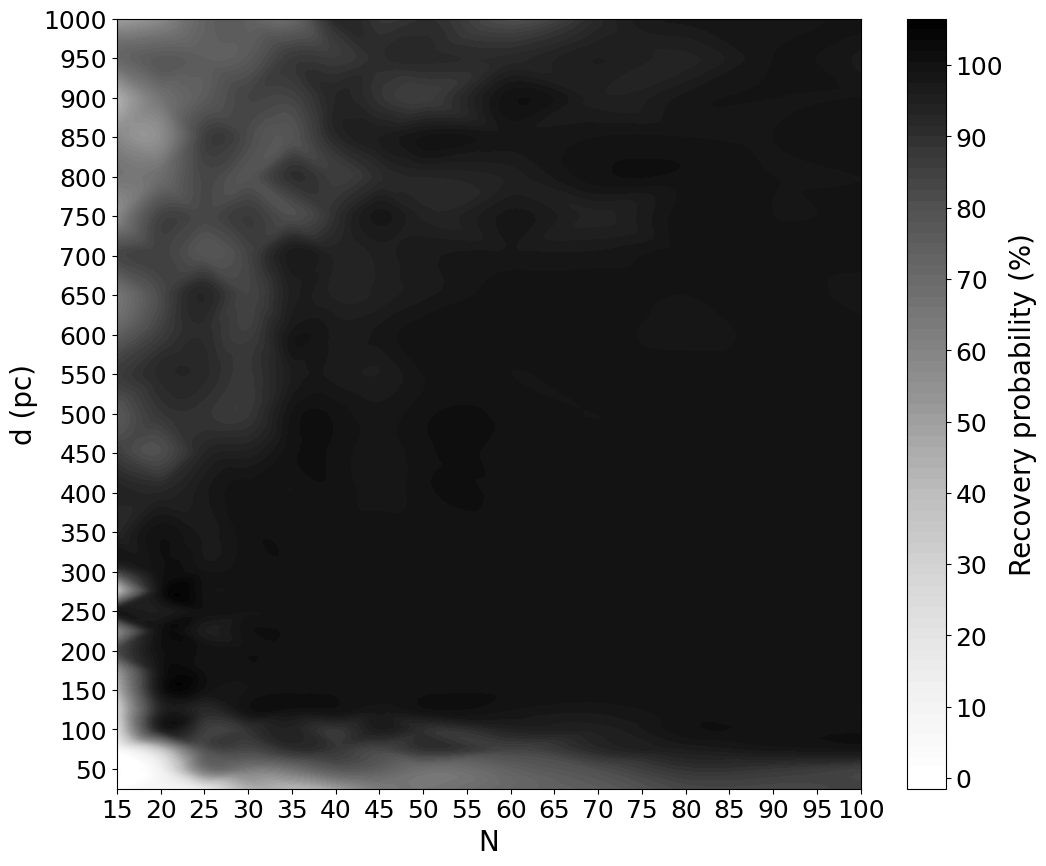}
    \caption{Recovery probability for the test association as a function of its number of OB members ($N$) and its central line-of-sight distance ($d$), as described in Section \ref{false_negatives}.}
    \label{Contours_FalseNegatives}
\end{figure}

The results of this process are shown in Fig. \ref{Contours_FalseNegatives}. As expected, the recovery probability of an OB association placed in a randomized field of synthetic OB stars increases with the number of members, and is at its lowest for the nearest and most distant associations with the smallest number of members. We already pointed out the limitations of our method for recovering smaller and more distant OB associations in \citet{Quintana2023}, and will have to account it when extending our all-sky census of OB associations to greater distances.

Given that these simulations show our incompleteness towards small and close OB associations, we decided to add back the four low-probability groups closer than 200 pc, increasing our number of candidate OB associations to 68. The final steps of our selection process are outlined in Section \ref{filtering_extension}.

\section{Correction for virtual expansion}
\label{correction_virtual}

In Section \ref{linearkinematicages}, we estimated linear kinematic ages by fitting a gradient between the Galactic coordinate of the OB association members and their proper motion in that dimension. Since our census includes very nearby OB associations, they can suffer from an effect known as "virtual expansion" whereby a non-expanding group of stars that is moving towards us can appear to be expanding when studied using only plane-of-the-sky motions \citep{Brown1997,vanleeuwen2009,CantatGaudin2019b}.

To correct for these effects, the first step consists of estimating a median RV for each OB association. To that end, we have crossmatched our 2551 OB association members with the following spectroscopic surveys (used in the following priority order):

\begin{itemize}
    \item Apache Point Observatory Galactic Evolution Experiment (APOGEE) DR17 spectroscopic sample from SDSS-V \citep{Garcia,Abdu}, to which we applied the criteria from  \citet{Cerqui2025} for selecting good RVs, i.e. S/N $>$ 50.
    \item \textit{Gaia} DR3 \citep{Katz2023}, to which we applied the correction $\rm R_V (corrected)= R_V - 0.002755 \, g^2_{rvs\_mag}+0.55863 \, g_{rvs\_mag}-2.81129$ from \citet{Katz2023}, as well as the magnitude-dependent correction $\rm R_V \, (corrected) = R_V - 7.98+1.135 \, g_{rvs\_mag} $ for stars with 8500 $\leq$ rv\_template\_teff $\leq$ 14,500 K and 6 $\leq$ grvs\_mag $\leq$ 12 mag from \citet{Blomme2023}.
    \item LAMOST DR6 low-resolution spectroscopic sub-survey \citep{Xiang2022LAMOST} specifically focused on the early-type stars \citep{Deng2012,Zhao2012,Liu2014}. We corrected RVs by an offset of 5 km s$^{-1}$ as in \citet{Katz2023}.
    \item Gaia-ESO survey \citep{Gilmore2022}, using the dedicated analysis of the spectra of hot stars \citep{Blomme2022}.
\end{itemize}

This process results in 1366 OB association members, hence $\sim$54 \% of our list, with a reliable RV measurement. This allowed us to estimate a median RV for each OB association. We have then used Eq. A3 from \citet{Brown1997} to estimate $V_{\rm sys}$ the radial component of the motion of the OB associations:

\begin{equation}
V_{\rm sys} = \frac{RV_{\rm med}}{\cos{(l_{\rm med}-l)} \, \cos{b_{\rm med}} \, cos{b} + \sin{b_{\rm med}} \, \sin{b}}
\end{equation}

\noindent where $l_{\rm med}$,  $b_{\rm med}$ and  $RV_{\rm med}$ are the median Galactic longitude, latitude and radial velocity of the OB association, respectively.

Once $V_{\rm sys}$ is computed, we were able to determine the proper motions due to virtual expansion, $\mu_{l, \rm virtual}$ and $\mu_{b, \rm virtual}$, also applying Eq. A3 from \citet{Brown1997}:

\begin{equation}
\mu_{l, \rm virtual} = \frac{V_{\rm sys}}{4.74 \, d} \, \cos{b_{\rm med}} \, \sin{(l_{\rm med}-l)}
\end{equation}
\begin{equation}
\mu_{b, \rm virtual} = \frac{V_{\rm sys}}{4.74 \, d} \, [\sin{b_{\rm med}} \cos{b} - \sin{b} \cos{b_{\rm med} \cos{(l_{\rm med}-l)}}]
\end{equation}

\noindent where $d$ stands for the SED-fitted distances of the OB association members, in units of kpc.

We then subtracted  $\mu_{l, \rm virtual}$ and $\mu_{b, \rm virtual}$ from the observed values of $\mu_l$ and $\mu_b$ of the OB association members in order to fit a linear gradient and estimate the kinematic ages of OB associations in Section \ref{linearkinematicages} that accounts for virtual expansion.

\section{Comparison with individual OB associations}
\label{indiv_OBassoc}

\begin{table*}
	\centering
	\caption{Comparison between our 2551 OB association members catalogues of individual OB associations. N$_{\rm C}$ is the total number of stars in the external catalogue within $\sqrt{X^2+Y^2} <$ 1 kpc whereas N$_{\rm O}$ stands for the number of our stars successfully crossmatched with the corresponding catalogue. 
    \label{CompIndivAssoc}}
	\renewcommand{\arraystretch}{1.3} 
	\begin{tabular}{lcccccr} 
		\hline
		Catalogue & Region & Data used & N$_{\rm C}$ & N$_{\rm O}$  \\
		\hline
        \citet{MiretRoig2022}  & Upper Scorpius \& Ophiuchus & \textit{Gaia} DR3 & 2816 & 15 \\
        \citet{Ratzenbock}  & Sco-Cen & \textit{Gaia} DR3 & 13,103 & 77 \\
        \citet{Sanchez2024} & Ori OB1 (Big Structures) & \textit{Gaia} DR3 & 5431 & 54 \\
        \citet{Cantat2019} &  Vela OB2  & \textit{Gaia} DR2 & 3527 & 27 \\
        \citet{Lim2022} & Mon OB1 & \textit{Gaia} DR2 & 726 & 15 \\
        \citet{Fleming2023} & Collinder 121 & \textit{Gaia} DR3 & 35 & 6 \\
        \citet{Szilagyi2023} & Cep OB2 & \textit{Gaia} DR3 & 874 & 20 \\
        \citet{Quintana2023} & Aur OB1 & \textit{Gaia} DR3 & 60 & 23 \\
		\hline
	\end{tabular}
\end{table*}

We have crossmatched our OB association members with recent, \textit{Gaia}-based catalogues of individual OB associations in Table \ref{CompIndivAssoc}, and detailed the comparison below:
\begin{itemize}
    \item \textbf{Upper Scorpius \& Ophiuchus:} \citet{MiretRoig2022} identified four stellar groups in Upper Scorpius, two in Ophiuchus and another young population, all of which belong to the Sco-Cen complex, and divided between 2816 stars. A crossmatch of our 103 stars in Sco-Cen gives 15 stars in common, including 6 in $\delta$ Sco, 4 in $\sigma$ Sco, 2 in $\pi$ Sco, 2 in $\alpha$ Sco and 1 in $\nu$ Sco. The dominating overlap is thus with Upper Scorpius, and all these stars are in Sco-Cen 1, which therefore includes notably this historical Sco-Cen subgroup.
    \item \textbf{Sco-Cen:} \citet{Ratzenbock} detected 37 groups split between 13,103 members. 77 out of our 103 OB association members in Sco-Cen, the vast majority of them, were successfully crossmatched with their catalogue. The breakdown gives 63 stars from Sco-Cen 1 (including 10 in $\sigma$ Cen, 10 in $\phi$ Lup, 9 in $\nu$ Cen, 7 in $\rho$ Sco, 6 in $\eta$ Lup and 5 in \textit{e} Lup) and 14 stars from Sco-Cen 2 (including 12 in V1062-Sco). Our Sco-Cen members are therefore split between many of their subgroups, an illustration of the rich structure unveiled in the individual studies of this complex. 
    \item \textbf{Ori OB1}: \citet{Sanchez2024} conducted a kinematical study of the Orion complex, wherein they identified many star clusters encompassed within Ori OB1, classified under two regimes. The first one, labelled as "Big Structures", consists of 13 groups previously discovered by \citet{Kounkel2018}, \citet{Kos2019}, \citet{Chen2020} and \citet{Swiggum2021}, and shared between 5431 members. A crossmatch between these stars and our 182 Ori OB1 members yields 54 stars in common. All of them belong to Ori OB1b, and are split between 17 stars in the ONC, 10 in OBP-Near, 6 in Ori-North, 5 in OBP-Far and OBP-b, 4 in $\sigma$ Ori and Briceño-1A, 3 in Briceño-1B and 1 in OBP-Far. From this comparison, it is clear that Ori OB1b provides a general overview of the Ori OB1 complex, as its strongest overlap is with the ONC which is by far the biggest cluster within \citep{Sanchez2024}. There is also no overlap with Ori OB1a, which appears as a smaller OB association in the region, closer to the Galactic Plane (see Fig. \ref{Map_OBAssociations}). 
    \item \textbf{Vela OB2}: \citet{Cantat2019} studied the 6D structure of the Vela OB2 complex, identifying 3527 likely members\footnote{While \citet{Armstrong2022} also performed a similar study but with \textit{Gaia} DR3 instead of DR2, they focused on low-mass stars, therefore there is no overlap with our OB association members.}. A crossmatch of our 156 members in Vela OB2 and Trumpler 10 gives 26 stars in common, 25 of which are in Vela OB2 and 1 in Tr. 10b. 
    \item \textbf{Mon OB1}: \citet{Lim2022} provided a \textit{Gaia} DR2 view of Mon OB1 and in doing so identified 728 Mon OB1 members, including 726 within $\sqrt{X^2+Y^2} <$ 1 kpc once crossmatched with \textit{Gaia} DR3. We have identified 15 out of our 64 Mon OB1 members in their catalogue, with a breakdown as follows. All but one are classified as type O or B (knowing their catalogue has 24 of such stars, the others are young stellar objects (YSOs) or pre-main sequence members), 9 are in the S Mon Group, 2 in Cone and 2 in THF15, whereas 1 is located in the halo of the Mon OB1 region.
    \item \textbf{Collinder 121}: \citet{Fleming2023} exploited \textit{Gaia} DR3 in order to revisit Collinder 121 and Canis Major OB2, and in doing so discovered a new OB associations situated at $\sim$800 pc and split between 4 subgroups. A crossmatch between their 35 members and our 205 members from Collinder 121 yields 6 stars in common, shared between their different subgroups. Given that Collinder 121 is our richest OB association of our catalogue, we argue that we have identified a larger complex in the region of Canis Major.
    \item \textbf{Cep OB2:} \citet{Szilagyi2023} identified 13 subgroups that belonged to Cep OB2. We have crossmatched our 63 Cep OB2 members and found 20 stars in common with theirs, including 13 in Cep OB2a and 7 in Cep OB2b. 9 of the crossmatching members from Cep OB2a are in Alessi-Teutsch 5 whilst all the crossmatching members from Cep OB2b are in NGC 7560, respectively aged 7.9 and 10.0 Myr, and located at the periphery and inside the Cepheus Bubble \citep{Szilagyi2023}. 
    \item \textbf{Aur OB1:} \citet{Quintana2023} identified 5 OB associations in the Auriga constellation. Because the closest one is the replacement for Aur OB1, we have crossmatched the 41 members of Aur OB1 with the 60 members of the first OB association from \citet{Quintana2023} within $\sqrt{X^2+Y^2} <$ 1 kpc. In doing so we have found 23 stars in common. Most of the first Auriga association in \citet{Quintana2023} lies beyond our coverage limit, at a median distance of $\sim$1.1 kpc, so we will only be able to obtain a full membership once we extend the all-sky census of OB associations.
\end{itemize}


\section{Missing historical OB associations}
\label{missing}

\begin{table*}

\centering
\caption{The missing historical OB associations from our list of 56 OB associations, where N and $d$ respectively stand for the median distance and the number of identified members within the provided reference. Reasons for exclusion are: a) Too extended in positional space (c.f. Section \ref{filtering_extension}), b) Lack of kinematic coherence in \textit{Gaia} DR3 data. \label{TableMissing}}

\begin{tabular}{ccccccc} 
\toprule
 Name & $d$ (pc) & N  & Data used & Reference & Reason for exclusion \\ 
\midrule
Cas-Tau & 125--300 & 83 & HIPPARCOS & \citet{HipparcosOBAssociations} & a, b \\
Monorion & 200 & 15  & HIPPARCOS & \citet{BouyAlves2015} & b \\
Vela OB5 & 250 & 24  & HIPPARCOS & \citet{BouyAlves2015} & b \\
Taurion & 320 & 36& HIPPARCOS & \citet{BouyAlves2015} & b  \\
Vul OB4 & 800 & 9 &  \textit{Gaia} DR2 & \citet{MelnikDambis2020} & b \\
Cyg OB4 & 800 & 2 &  \textit{Gaia} DR2 & \citet{MelnikDambis2020} & b \\
Cas OB14 & 880 & 8 &  \textit{Gaia} DR2 & \citet{MelnikDambis2020} & b \\
\hline

\end{tabular}
\end{table*}

In Section \ref{identification}, we emphasized that our final list of OB associations corresponds to a high-confidence sample, and that some of the initial groups we have rejected could still correspond to existing OB associations. We have listed these missing OB associations in Table \ref{TableMissing}, indicated whether some of them were found in our rejected groups (and, if so, with which degree of overlap), and the reasons for their exclusion from our final list.

\section{Comparison with star clusters}
\label{compoc}

In Table \ref{TableOC} we list the overlap between each of our OB associations and the related OCs from \citet{HuntReffert2024}, where we have also indicated the median age derived from the related OCs (displaying the range when there was overlap with several OCs), as explained in Section \ref{starclusters}.

\begin{table*}

\centering
\caption{The 56 OB associations with their related OCs from \citet{HuntReffert2024}, ordered from largest to smallest overlap of members. N$_{\rm OC}$ stands for the number of OB association members within OCs, and the related OCs age for the interval of ages from the crossmatching OCs.} \label{TableOC}
\begin{tabular}{ccccccc} 
\toprule
 ID &      Name &   N &  N$_{\rm OC}$ &  Fraction (\%) &  Related OCs age (Myr) &                                        Related OCs \\
\midrule
1 & Per OB3 & 15 & 11 & 73.3 & 122 & Melotte\_22 \\
2 & Sco-Cen 1 & 87 & 10 & 11.5 & 5--10 & HSC\_2733, OCSN\_96, OCSN\_92, HSC\_2468, HSC\_2907 \\
3 & Sco-Cen 2 & 16 & 9 & 56.2 & 10 & UPK\_640 \\
4 & Cep OB6 & 18 & 4 & 22.2 & 57--89 & Delta\_Cephei\_Cluster, Theia\_117, ADS\_16795 \\
5 & Ori OB1a & 26 & 4 & 15.4 & 13 & NGC\_2232 \\
6 & Her OB1a & 89 & 18 & 20.2 & 22--168 & RSG\_5, Roslund\_6, CWNU\_519,  \\
 &  &  &  &  & &  CWNU\_1032, Theia\_96, CWNU\_1075 \\
7 & Per OB2 & 34 & 7 & 20.6 & 5--17 & HSC\_1262, OC\_0279, Alessi-Teutsch\_10 \\
8 & Her OB1b & 32 & 5 & 15.6 & 27 & Stephenson\_1 \\
9 & Ori OB1b & 156 & 31 & 19.9 & 4--13 & ASCC\_20, UBC\_17a, Sigma\_Orionis, NGC\_1980, OCSN\_61, \\
 &  &  & &  & &  ASCC\_21, Briceno\_1, ASCC\_19, Theia\_13, ASCC\_18 \\
10 & Vela OB2 & 40 & 9 & 22.5 & 7--9 & OC\_0470, Pozzo\_1, OC\_0479 \\
11 & Tr. 10a & 51 & 22 & 43.1 & 37--56 & Alessi\_5, BH\_99, OCSN\_89, Theia\_246 \\
12 & Tr. 10b & 39 & 32 & 82.1 & 125 & NGC\_2516 \\
13 & Cas OB15 & 66 & 27 & 40.9 & 7--146 & RSG\_8, Stock\_12, Alessi\_20, NGC\_7429, \\
 &  &  &  & &  & ASCC\_127, Theia\_391, OCSN\_41, OCSN\_40, HSC\_899 \\
14 & Tr. 10c & 26 & 21 & 80.8 & 37 & Trumpler\_10 \\
15 & CMa OB3 & 33 & 24 & 72.7 & 109--132 & NGC\_2422, OCSN\_77 \\
16 & Cyg OB10a & 62 & 31 & 50.0 & 6--104 & Theia\_38, UBC\_26, UPK\_82, Theia\_836, ASCC\_105, UPK\_72 \\
17 & Cyg OB10b & 18 & 6 & 33.3 & 122 & Alessi\_12 \\
18 & Cyg OB10c & 39 & 13 & 33.3 & 57--110 & Roslund\_5, HSC\_601, HSC\_633 \\
19 & Sco OB3 & 38 & 12 & 31.6 & 100--128 & Theia\_181, Alessi\_24, UBC\_11 \\
20 & Car OB3a & 17 & 14 & 82.4 & 57 & HSC\_2384 \\
21 & Tau OB1 & 23 & 20 & 87.0 & 117 & NGC\_1647 \\
22 & Lac OB1 & 18 & 11 & 61.1 & 28 & UPK\_168 \\
23 & Car OB3b & 40 & 9 & 22.5 & 48--162 & Theia\_1170, HSC\_2271, Theia\_149, HSC\_2298 \\
24 & Sco OB5 & 22 & 18 & 81.8 & 78 & NGC\_6124 \\
25 & Sgr OB2a & 63 & 52 & 82.5 & 96--123 & IC\_4725, Alessi\_40, HSC\_181 \\
26 & Cam OB2a & 89 & 31 & 34.8 & 38--114 & Trumpler\_3, Stock\_7, Stock\_23, Theia\_85, HSC\_1080, Theia\_654 \\
27 & Aql OB2 & 38 & 14 & 36.8 & 36--123 & CWNU\_1074, OC\_0069, CWNU\_313, ASCC\_106 \\
28 & Sco OB6 & 28 & 5 & 17.9 & 148 & HSC\_2865, Theia\_551 \\
29 & Cam OB2b & 25 & 12 & 48.0 & 90 & Trumpler\_2 \\
30 & Sgr OB2b & 26 & 26 & 100.0 & 84 & Collinder\_394 \\
31 & Mon OB1 & 64 & 14 & 21.9 & 5--358 & NGC\_2264, Theia\_282, CWNU\_418\\
 &  &  &  &  & & NGC\_2202, Mon\_OB1-D, HSC\_1666 \\
32 & Cep OB7 & 25 & 15 & 60.0 & 16--125 & Alessi\_37, UPK\_166 \\
33 & Cir OB2 & 58 & 21 & 36.2 & 82--232 & NGC\_5662, OC\_0588, Loden\_1194 \\
34 & CMa OB4 & 25 & 21 & 84.0 & 154 & NGC\_2287 \\
35 & Cyg OB7 & 29 & 8 & 27.6 & 4--25 & NGC\_7039, UPK\_127, Theia\_26, HSC\_705, HSC\_699 \\
36 & Vela OB3 & 22 & 14 & 63.6 & 125--313 & Theia\_274, HSC\_2106, HSC\_2058 \\
37 & Cir OB3a & 30 & 20 & 66.7 & 115 & NGC\_6025 \\
38 & Collinder 121 & 205 & 39 & 19.0 & 8--282 & HSC\_1865, OC\_0395, OC\_0401, Collinder\_132, \\
& &  & &  &  &  Alessi\_33, OC\_0407, HSC\_1936, HSC\_1894, \\
& &  & &  &  &  Gulliver\_21, CWNU\_45, CWNU\_340, HSC\_1913, HSC\_1911 \\
39 & Sge OB1 & 46 & 6 & 13.0 & 63--77 & UPK\_56, UPK\_54 \\
40 & Cir OB3b & 26 & 5 & 19.2 & 5 & ASCC\_79 \\
41 & Gem OB3 & 109 & 49 & 45.0 & 67--202 & NGC\_2168, COIN-Gaia\_23, UPK\_381, HSC\_1467 \\
42 & Mon OB5 & 41 & 18 & 43.9 & 120--139 & NGC\_2301, HSC\_1635, ASCC\_29 \\
43 & Cep OB3 & 35 & 9 & 25.7 & 4--5 & Theia\_4, FSR\_0416, UBC\_178, HSC\_835, ASCC\_125 \\
44 & Sct OB4 & 47 & 3 & 6.4 & 33--47 & UBC\_1595, HSC\_330, HSC\_276 \\
45 & Cam OB1 & 57 & 39 & 68.4 & 5--146 & Collinder\_463, FSR\_0569, UBC\_415, CWNU\_77 \\
46 & Cep OB8 & 34 & 21 & 61.8 & 106 & NGC\_7243 \\
47 & Ara OB1 & 87 & 32 & 36.8 & 8--187 & ASCC\_85, NGC\_6178, NGC\_6250, HXHWL\_10, \\
& & & & & & HSC\_2802, UPK\_630, HSC\_2789 \\
48 & Cep OB2a & 41 & 20 & 48.8 & 6 & OC\_0185, Alessi-Teutsch\_5, Pismis-Moreno\_1 \\
49 & Cep OB2b & 22 & 9 & 40.9 & 9--57 & NGC\_7160, UBC\_10b \\
50 & Sge OB2 & 48 & 10 & 20.8 & 24--169 & ASCC\_107, OC\_0077, Theia\_316, Roslund\_1, HSC\_447 \\
51 & Vela OB4 & 30 & 22 & 73.3 & 165 & NGC\_2546, CWNU\_435 \\
52 & Vela OB6 & 46 & 26 & 56.5 & 4--8 & Alessi\_43, BH\_56, Pismis\_5, Collinder\_197, OC\_0467, FSR\_1421 \\
53 & Cir OB4 & 35 & 16 & 45.7 & 91 & NGC\_6087 \\
54 & Mon OB6 & 29 & 12 & 41.4 & 144 & NGC\_2323 \\
55 & Car OB4 & 45 & 25 & 55.6 & 166 & NGC\_3114 \\
56 & Aur OB1 & 41 & 3 & 7.3 & 95 & CWNU\_518 \\
\bottomrule
\end{tabular}
\end{table*}

\section{Comparison with star clusters, co-moving groups and structures}
\label{compgroups}

In Table \ref{TableGroups} we list the overlap between each of our OB associations and the related star clusters, co-moving groups and structures from \citet{Kounkel2020}. Just like for Table \ref{TableOC}, we have indicated the median age derived from the related OCs and displaying the range in cases of several overlapping OCs. 

\begin{table*}

\centering
\caption{The 56 OB associations with their related star clusters, co-moving groups and structures from \citet{Kounkel2020}, ordered from largest to smallest overlap of members, labeling the groups by their ID in case they did not have a recorded name. N$_{\rm G}$ stands for the number of OB association members within these groups. Just like for Table \ref{TableOC}, we have displayed the range of ages from the related groups in case our OB associations were crossmatching with several ones. \label{TableGroups}}

\begin{tabular}{ccccccccccc} 
\toprule
 ID & Name & N & N$_{\rm G}$ & Fraction (\%) & Related groups age (Myr) & Related groups \\ 
\midrule
1 & Per OB3 & 15 & 8 & 53.3 & 182 & Pleiades \\
2 & Sco-Cen 1 & 87 & 20 & 23.0 & 7--13 & CrA, Upper Sco \\
3 & Sco-Cen 2 & 16 & 11 & 68.8 & 7--13 & CrA, Upper Sco \\
4 & Cep OB6 & 18 & 6 & 33.3 & 44--71 & Alpha per, Cep OB6, ASCC 123 \\
5 & Ori OB1a & 26 & 12 & 46.2 & 5--22 & NGC 2232, Orion \\
6 & Her OB1a & 89 & 28 & 31.5 & 34--65 & Roslund 5, RSG 5, Group 98 \\
7 & Per OB2 & 34 & 19 & 55.9 & 4--8 & Per OB2, NGC 1333, IC 348 \\
8 & Her OB1b & 32 & 22 & 68.8 & 22 & Stephenson 1 \\
9 & Ori OB1b & 156 & 119 & 76.9 & 5 & Orion \\
10 & Vela OB2 & 40 & 22 & 55.0 & 12--37 & Vela OB2, Group 120 \\
11 & Tr. 10a & 51 & 22 & 43.1 & 58--95 & Trumpler 10, Group 246, Group 149 \\
12 & Tr. 10b & 39 & 37 & 94.9 & 145 & NGC 2516 \\
13 & Cas OB15 & 66 & 34 & 51.5 & 13--166 & LDN 1219, Stock 12, Alessi 20, ASCC 128 \\
14 & Tr. 10c & 26 & 12 & 46.2 & 58 & Trumpler 10 \\
15 & CMa OB3 & 33 & 26 & 78.8 & 105--178 & NGC 2422, Group 379 \\
16 & Cyg OB10a & 62 & 34 & 54.8 & 8--316 & Group 38, Group 148, Group 836, Roslund 5, ASCC 105, Group 142 \\
17 & Cyg OB10b & 18 & 13 & 72.2 & 98 & Alessi 12 \\
18 & Cyg OB10c & 39 & 10 & 25.6 & 49--65 & Roslund 5, Group 98 \\
19 & Sco OB3 & 38 & 14 & 36.8 & 25--145 & Group 181, Alessi 24, Group 447, BH 164 \\
20 & Car OB3a & 17 & 11 & 64.7 & 85 & NGC 2925 \\
21 & Tau OB1 & 23 & 20 & 87.0 & 251--263 & NGC 1647, Group 741 \\
22 & Lac OB1 & 18 & 8 & 44.4 & 35 & UBC165 \\
23 & Car OB3b & 40 & 5 & 12.5 & 66 & Group 149 \\
24 & Sco OB5 & 22 & 17 & 77.3 & 123 & NGC 6124 \\
25 & Sgr OB2a & 63 & 30 & 47.6 & 178--251 & IC 4725, ASCC 97 \\
26 & Cam OB2a & 89 & 25 & 28.1 & 74--224 & Trumpler 3, Stock 7, Group 574, Group 752, Group 384 \\
27 & Aql OB2 & 38 & 17 & 44.7 & 74 & Alessi 44 \\
28 & Sco OB6 & 28 & 2 & 7.1 & 36--195 & Group 551, NGC 6178 \\
29 & Cam OB2b & 25 & 10 & 40.0 & 182 & Trumpler 2 \\
30 & Sgr OB2b & 26 & 18 & 69.2 & 110 & Collinder 394 \\
31 & Mon OB1 & 64 & 22 & 34.4 & 3--123 & NGC 2264, Group 189, NGC 2169, Group 261 \\
32 & Cep OB7 & 25 & 14 & 56.0 & 35--145 & Alessi 37, UBC165 \\
33 & Cir OB2 & 58 & 21 & 36.2 & 63--195 & NGC 5662, Group 857, Group 350, Loden 1194 \\
34 & CMa OB4 & 25 & 20 & 80.0 & 102 & NGC 2287 \\
35 & Cyg OB7 & 29 & 13 & 44.8 & 2--129 & NGC 7039, Group 563, Group 3 \\
36 & Vela OB3 & 22 & 3 & 13.6 & 79 & Group 274 \\
37 & Cir OB3a & 30 & 20 & 66.7 & 100 & NGC 6025 \\
38 & Collinder 121 & 205 & 33 & 16.1 & 15--112 & Group 87, Group 86, Collinder 132, NGC 2362, Gulliver 21, Group 283 \\
39 & Sge OB1 & 46 & 6 & 13.0 & 85--123 & Group 495, Group 268, Group 2312 \\
40 & Cir OB3b & 26 & 16 & 61.5 & 9--120 & ASCC 79, Group 353 \\
41 & Gem OB3 & 109 & 50 & 45.9 & 52--229 & NGC 2168, COIN-Gaia 23, Group 859, Group 760 \\
42 & Mon OB5 & 41 & 14 & 34.1 & 115 & NGC 2323, Group 2583 \\
43 & Cep OB3 & 35 & 9 & 25.7 & 5--26 & Group 4, Cep OB3b, Group 1788 \\
44 & Sct OB4 & 47 & 7 & 14.9 & 59 & UBC116 \\
45 & Cam OB1 & 57 & 39 & 68.4 & 10--200 & Collinder 463, UBC187 \\
46 & Cep OB8 & 34 & 21 & 61.8 & 135--148 & NGC 7243, ASCC 115 \\
47 & Ara OB1 & 87 & 33 & 37.9 & 32--110 & NGC 6178, NGC 6250, ASCC 85, Harvard 10, UBC319, UBC544 \\
48 & Cep OB2a & 41 & 20 & 48.8 & 5--22 & IC 1396, Pismis Moreno 1 \\
49 & Cep OB2b & 22 & 14 & 63.6 & 17 & NGC 7160 \\
50 & Sge OB2 & 48 & 4 & 8.3 & 20--43 & ASCC 107, Group 1744 \\
51 & Vela OB4 & 30 & 8 & 26.7 & 102--123 & NGC 2546, Group 279 \\
52 & Vela OB6 & 46 & 17 & 37.0 & 5 & Collinder 197 \\
53 & Cir OB4 & 35 & 15 & 42.9 & 74 & NGC 6087 \\
54 & Mon OB6 & 29 & 10 & 34.5 & 115 & NGC 2323 \\
55 & Car OB4 & 45 & 32 & 71.1 & 85 & NGC 3114 \\
56 & Aur OB1 & 41 & 4 & 9.8 & 58--209 & ASCC 12, Group 2752, RSG 1 \\
\bottomrule
\end{tabular}
\end{table*}


\bsp	
\label{lastpage}
\end{document}